\documentclass[12pt]{article}
\usepackage{amsmath}
  \usepackage{graphicx,xcolor,float}
\usepackage{indentfirst}
\usepackage{multirow}
\usepackage{textcomp}
\usepackage{hyperref}
\usepackage{amssymb}
\usepackage{amsthm}
\usepackage{overpic}
\usepackage{bbm,bm,epsfig,ulem}
\usepackage{url}
\usepackage{caption}
\usepackage{subfigure}
\usepackage{cite}

\def\thefootnote{\fnsymbol{footnote}}

\begin{document}

\begin{center}
{\Large\bf 3D mapping of the effective Majorana neutrino masses \\ with
neutrino oscillation data}
\end{center}

\vspace{0.2cm}

\begin{center}
{\bf Ce-ran Hu$^{1}$}
\footnote{E-mail: huceran18@mails.ucas.ac.cn},
{\bf Zhi-zhong Xing$^{1,2,3}$}
\footnote{E-mail: xingzz@ihep.ac.cn}
\\
{\small $^{1}$School of Physical Sciences,
University of Chinese Academy of Sciences, Beijing 100049, China \\
$^{2}$Institute of High Energy Physics, Chinese Academy of Sciences, Beijing 100049, China \\
$^{3}$Center of High Energy Physics, Peking University, Beijing 100871, China}
\end{center}

\vspace{2cm}
\begin{abstract}
With the help of current neutrino oscillation data, we illustrate the
three-dimensional (3D) profiles of all the six distinct effective Majorana
neutrino masses $|\langle m\rangle^{}_{\alpha \beta}|$ (for $\alpha, \beta
= e, \mu, \tau$) with respect to the unknown neutrino mass scale and effective Majorana
CP phases. Some salient features of $|\langle m\rangle^{}_{\alpha \beta}|$ and their
phenomenological implications are discussed in both the normal and inverted
neutrino mass ordering cases.
\end{abstract}

\newpage

\def\thefootnote{\arabic{footnote}}
\setcounter{footnote}{0}
\setcounter{figure}{0}

\section{Introduction}

In the standard three-flavor scheme the phenomenology of neutrino oscillations can be
fully described in terms of the neutrino mass-squared differences $\Delta m^2_{ij}
\equiv m^2_i - m^2_j$ (for $i, j = 1, 2, 3$), the lepton flavor mixing angles
$\theta^{}_{ij}$ (for $ij = 12, 13, 23$) and the Dirac
CP-violating phase $\delta$~\cite{Zyla:2020zbs}.
A series of successful atmospheric, solar, reactor and accelerator neutrino (or
antineutrino) oscillation experiments have allowed us to determine the values of
$\Delta m^2_{21}$, $|\Delta m^2_{31}|$ (or $|\Delta m^2_{32}|$), $\theta^{}_{12}$,
$\theta^{}_{13}$ and $\theta^{}_{23}$ to a quite high degree of accuracy~\cite{Zyla:2020zbs},
but our present knowledge on the sign of $\Delta m^2_{31}$
(or $\Delta m^2_{32}$)~\cite{Capozzi:2017ipn,Esteban:2020cvm}
and the value of $\delta$~\cite{Abe:2019vii}
remains rather poor, at most at the $3\sigma$ level. The next-generation
reactor and accelerator neutrino (or antineutrino) oscillation experiments are
expected to resolve these two important issues in the near future.

Once all the neutrino oscillation parameters are well measured, to what extent can
one pin down the flavor structure of three neutrino species? The answer
to this question is certainly dependent on the nature of massive neutrinos. Here let us
assume that each massive neutrino is a Majorana fermion; i.e., it is its own
antiparticle~\cite{Majorana:1937vz}. Then we are left with three unknown
parameters which are completely
insensitive to normal neutrino oscillation experiments: the absolute neutrino mass
scale and two Majorana CP-violating phases. These three free parameters can only be
determined or constrained with the help of some {\it non-oscillation} processes
\footnote{Of course, both the absolute neutrino mass scale and the Majorana CP
phases are sensitive to the lepton-number-violating neutrino-antineutrino
oscillations~\cite{Pontecorvo:1957cp}, but the latter cannot be observed in any
realistic experiments simply because their probabilities are typically suppressed
by the tiny factors $m^2_i/E^2$ with $m^{}_i$ being the neutrino mass and $E$
being the neutrino beam energy~\cite{Schechter:1980gk,deGouvea:2002gf,Xing:2013ty}.
For instance, $m^2_i/E^2 \lesssim {\cal O}(10^{-14})$ is expected for the reactor
antineutrino beam with $E \sim {\cal O}(1)$ MeV and $m^{}_i \lesssim 0.1$ eV.},
such as the beta decays, neutrinoless double-beta ($0\nu 2\beta$) decays and
cosmological observations~\cite{Zyla:2020zbs}. In particular, the Majorana CP
phases are only sensitive to those lepton-number-violating processes
which have never been measured~\cite{Rodejohann:2011mu}. Before such challenging
measurements are successfully implemented in the foreseeable future, one has to
use the available neutrino oscillation data to constrain the moduli of six
distinct effective Majorana neutrino masses
\begin{eqnarray}
\langle m\rangle^{}_{\alpha\beta} \equiv \sum^3_{i=1} m^{}_i U^{}_{\alpha i}
U^{}_{\beta i} \;
\end{eqnarray}
in the basis of {\it diagonal} charged-lepton flavors,
where $U^{}_{\alpha i}$ (for $\alpha = e, \mu, \tau$ and $i = 1, 2, 3$) are the
elements of the Pontecorvo-Maki-Nakawaga-Sakata (PMNS) lepton flavor mixing
matrix~\cite{Pontecorvo:1957cp,Pontecorvo:1967fh,Maki:1962mu}. This is actually
the only {\it model-independent} way to reconstruct the effective Majorana neutrino
mass matrix at low energies,
\begin{eqnarray}
M^{}_{\nu} = U D^{}_{\nu} U^T = \begin{pmatrix} \langle m\rangle^{}_{ee} &
\langle m\rangle^{}_{e\mu} & \langle m\rangle^{}_{e\tau} \cr
\langle m\rangle^{}_{e\mu} & \langle m\rangle^{}_{\mu\mu} &
\langle m\rangle^{}_{\mu\tau} \cr
\langle m\rangle^{}_{e\tau} & \langle m\rangle^{}_{\mu\tau} &
\langle m\rangle^{}_{\tau\tau} \cr \end{pmatrix} \; ,
\end{eqnarray}
where $D^{}_{\nu} = {\rm Diag}\{m^{}_{1}, m^{}_{2}, m^{}_{3}\}$. The texture of
$M^{}_\nu$ may help indicate some kind of underlying flavor symmetry, which will be
greatly useful for building a predictive neutrino mass
model~\cite{Xing:2019vks,Feruglio:2021sir}.

Given the available neutrino oscillation data, one has so far paid a lot of attention
to the two-dimensional (2D) mapping of $|\langle m\rangle^{}_{\alpha\beta}|$ as functions
of the smallest neutrino mass $m^{}_1$ (normal mass ordering, NMO) or $m^{}_3$ (inverted
mass ordering, IMO) ~\cite{Xing:2019vks}, especially to the allowed region of
$|\langle m\rangle^{}_{ee}|$ which is closely associated with the observability of
the $0\nu 2\beta$ decays. In such a 2D mapping one usually requires the unknown
CP-violating phases of $U$ to vary from $0$ to $2\pi$, and hence it is almost
impossible to see the explicit dependence of $|\langle m\rangle^{}_{\alpha\beta}|$
on those phase parameters. A three-dimensional (3D) mapping of
$|\langle m\rangle^{}_{ee}|$ has recently been done in order to better understand
its parameter space as constrained by a large or complete cancellation in the NMO
case~\cite{Xing:2015zha,Xing:2016ymd,Cao:2019hli}. It is proved that the 3D mapping
approach does have its remarkable merits in revealing the salient features of
the effective Majorana neutrino mass term $|\langle m\rangle^{}_{ee}|$, especially
in the situation of no experimental information about the Majorana CP phases of $U$.

The present work aims to map the 3D profiles of all the six effective Majorana neutrino
masses $|\langle m\rangle^{}_{\alpha\beta}|$ with the help of current neutrino
oscillation data. We are motivated by the fact that it is impossible to determine
the Majorana CP phases only from a measurement of $|\langle m\rangle^{}_{ee}|$ in
the future $0\nu 2\beta$-decay experiments \cite{Barger:2002vy}. In other words,
one has to make every effort to go beyond the $0\nu 2\beta$ decays and probe
other possible lepton-number-violating processes after the Majorana nature of
massive neutrinos is established from a successful $0\nu 2\beta$-decay experiment.
The ultimate goal of such efforts is to pin down all the CP-violating phases of
$U$ and have a full understanding of the flavor structure of massive neutrinos.
Although this goal seems to be too remote from our today's experimental techniques,
the 3D mapping of $|\langle m\rangle^{}_{\alpha\beta}|$ may at least help
shed light on the underlying flavor symmetry behind the observed neutrino mass
spectrum and flavor mixing pattern. Our study in this connection is therefore
meaningful and useful.

The remaining parts of this paper are organized as follows. In section 2 we define
the effective Majorana phases $\rho$ and $\sigma$, which are associated respectively
with $m^{}_1$ and $m^{}_2$, for each of $|\langle m\rangle^{}_{\alpha\beta}|$.
Then we determine the upper and lower bounds of $|\langle m\rangle^{}_{\alpha\beta}|$
with respect to $\sigma$. Section 3 is devoted to the 3D mapping of
$|\langle m\rangle^{}_{\alpha\beta}|$ by inputting the best-fit values of
$\Delta m^2_{21}$, $|\Delta m^2_{31}|$ (or $|\Delta m^2_{32}|$),
$\theta^{}_{12}$, $\theta^{}_{13}$, $\theta^{}_{23}$ and $\delta$ that have
been extracted from a global analysis of current neutrino oscillation data.
The salient features of six $|\langle m\rangle^{}_{\alpha\beta}|$ and their
respective phenomenological implications are discussed in both the NMO and IMO
cases. Section 4 is devoted to some further discussions about the
4D plots of $|\langle m\rangle^{}_{\alpha\beta}|$, about the dependence
of the 3D profiles of $|\langle m\rangle^{}_{\alpha\beta}|$ on the redefinition
of their effective Majorana phases, and about an approximate $\mu$-$\tau$ symmetry
and (or) texture zeros that have emerged in the numerical results of $|\langle m\rangle^{}_{\alpha\beta}|$. We summarize our main points in section 5.

\section{Expressions of $|\langle m\rangle^{}_{\alpha\beta}|$ and their extrema}

Let us adopt the standard parametrization of the $3\times 3$ PMNS lepton flavor mixing
matrix $U$ as advocated by the Particle Data Group~\cite{Zyla:2020zbs}:
\begin{eqnarray}
U = \begin{pmatrix}
c^{}_{12} c^{}_{13} & s^{}_{12} c^{}_{13} &
s^{}_{13} e^{-{\rm i} \delta} \cr
-s^{}_{12} c^{}_{23} - c^{}_{12}
s^{}_{13} s^{}_{23} e^{{\rm i} \delta} & c^{}_{12} c^{}_{23} -
s^{}_{12} s^{}_{13} s^{}_{23} e^{{\rm i} \delta} & c^{}_{13}
s^{}_{23} \cr
s^{}_{12} s^{}_{23} - c^{}_{12} s^{}_{13} c^{}_{23}
e^{{\rm i} \delta} &- c^{}_{12} s^{}_{23} - s^{}_{12} s^{}_{13}
c^{}_{23} e^{{\rm i} \delta} &  c^{}_{13} c^{}_{23} \cr
\end{pmatrix}
P^{}_\nu \; ,
\end{eqnarray}
where $c^{}_{ij} \equiv \cos\theta^{}_{ij}$ and $s^{}_{ij} \equiv \sin\theta^{}_{ij}$
(for $ij = 12, 13, 23$), $\delta$ stands for the Dirac CP phase,
and $P^{}_\nu = {\rm Diag}\{e^{{\rm i} \xi_{1}/2}, e^{{\rm i} \xi_{2}/2}, 1\}$ contains
two Majorana CP phases which are insensitive to normal neutrino oscillations.
Given the basis in which the flavor eigenstates of three charged leptons are
identical with their mass eigenstates, one may reconstruct the effective Majorana
neutrino masses $\langle m\rangle^{}_{\alpha\beta}$ by means of Eqs.~(2) and (3).
The explicit results are
\begin{eqnarray}
&& \langle m \rangle^{}_{ee} =
m^{}_{1} c^2_{12} c^2_{13} e^{{\rm i} \xi^{}_{1}} + m^{}_{2} s^2_{12} c^2_{13} e^{{\rm i}\xi^{}_{2}} + m^{}_{3} s^2_{13} e^{-2{\rm i} \delta} \; ,
\nonumber \\
&& \langle m \rangle^{}_{\mu\mu} = m^{}_{1} \left(s^{}_{12} c^{}_{23} + c^{}_{12} s^{}_{23} s^{}_{13} e^{{\rm i}\delta}\right)^{2} e^{{\rm i}\xi^{}_{1}}
+ m^{}_{2}\left(c^{}_{12} c^{}_{23} - s^{}_{12} s^{}_{23} s^{}_{13} e^{{\rm i}
\delta}\right)^{2} e^{{\rm i}\xi^{}_{2}} + m^{}_{3} c^2_{13} s^2_{23} \; ,
\nonumber \\
&& \langle m \rangle^{}_{\tau\tau }= m^{}_{1}\left(s^{}_{12} s^{}_{23} - c^{}_{12} c^{}_{23} s^{}_{13} e^{{\rm i}\delta}\right)^{2} e^{{\rm i}\xi^{}_{1}} + m^{}_{2}\left(c^{}_{12} s^{}_{23} + s^{}_{12} c^{}_{23} s^{}_{13} e^{{\rm i}\delta}\right)^{2} e^{{\rm i}\xi^{}_{2}} + m^{}_{3} c^2_{13} c^2_{23} \; , \hspace{1cm}
\end{eqnarray}
and
\begin{eqnarray}
&&\langle m \rangle^{}_{e\mu} = -m^{}_{1} c^{}_{12} c^{}_{13}\left(s^{}_{12} c^{}_{23} +
c^{}_{12} s^{}_{23} s^{}_{13} e^{{\rm i}\delta}\right) e^{{\rm i}\xi^{}_{1}} + m^{}_{2} s^{}_{12} c^{}_{13}\left(c^{}_{12} c^{}_{23} - s^{}_{12} s^{}_{23} s^{}_{13} e^{{\rm i}\delta}\right)
e^{{\rm i}\xi^{}_{2}} \hspace{1.2cm}
\nonumber \\
&& \hspace{1.6cm} +m^{}_{3} c^{}_{13} s^{}_{23} s^{}_{13} e^{-{\rm i}\delta} \; ,
\nonumber \\
&&\langle m \rangle^{}_{e\tau} = +m^{}_{1} c^{}_{12} c^{}_{13}\left(s^{}_{12} s^{}_{23} -
c^{}_{12} c^{}_{23} s^{}_{13} e^{{\rm i}\delta}\right) e^{{\rm i}\xi^{}_{1}} - m^{}_{2} s^{}_{12} c^{}_{13}\left(c^{}_{12} s^{}_{23} + s^{}_{12} c^{}_{23} s^{}_{13} e^{{\rm i}\delta}\right)
e^{{\rm i}\xi^{}_{2}}
\nonumber \\
&& \hspace{1.6cm} +m^{}_{3} c^{}_{13} c^{}_{23} s^{}_{13} e^{-{\rm i}\delta} \; ,
\nonumber \\
&&\langle m \rangle^{}_{\mu\tau} = -m^{}_{1}\left(s^{}_{12} c^{}_{23} + c^{}_{12} s^{}_{23}
s^{}_{13} e^{{\rm i}\delta}\right)\left(s^{}_{12} s^{}_{23} - c^{}_{12} c^{}_{23} s^{}_{13}
e^{{\rm i}\delta}\right) e^{{\rm i}\xi^{}_{1}}
\nonumber \\
&& \hspace{1.6cm} -m^{}_{2}\left(c^{}_{12} c^{}_{23} - s^{}_{12} s^{}_{23} s^{}_{13} e^{{\rm i}\delta}\right)\left(c^{}_{12} s^{}_{23} + s^{}_{12} c^{}_{23} s^{}_{13} e^{{\rm i}\delta}\right)
e^{{\rm i}\xi^{}_{2}} + m^{}_{3} c^2_{13} c^{}_{23} s^{}_{23} \; .
\end{eqnarray}
Except $|\langle m\rangle^{}_{ee}|$, the moduli of the other five effective Majorana
neutrino masses are all dependent on the CP phase $\delta$. Hence $\delta$ is also
of the Majorana nature, although it determines the strength of CP violation in normal
neutrino oscillations and often referred to as the ``Dirac" CP phase.

\subsection{The effective Majorana phases}

To effectively map the profile of $|\langle m\rangle^{}_{\alpha\beta}|$ with the help of
current neutrino oscillation data, we choose to redefine the phases of three
components of $|\langle m\rangle^{}_{\alpha\beta}|$ proportional to $m^{}_i$ (for
$i=1,2,3$) such that the new expression of $|\langle m\rangle^{}_{\alpha\beta}|$ depends
only on two {\it effective} phase parameters $\rho$ and $\sigma$. Of course, these
two phases must be some simple functions of the original CP-violating phases of $U$ (i.e.,
$\delta$, $\xi^{}_1$ and $\xi^{}_2$). For each of the six effective Majorana neutrino
masses, the corresponding definition of $\rho$ and $\sigma$ is given as follows.
Let us assign the effective phases
$\rho$ and $\sigma$ to the terms of $|\langle m\rangle^{}_{\alpha\beta}|$ that are
proportional to $m^{}_1$ and $m^{}_2$, respectively
\footnote{Note that we should have defined $\rho^{}_{\alpha\beta}$ and $\sigma^{}_{\alpha\beta}$
for six different $|\langle m\rangle^{}_{\alpha\beta}|$. But for the sake of simplicity, here
we have omitted such subscripts for the two effective phase parameters. Hence one should keep
in mind that the meanings and values of $\rho$ and $\sigma$ for different
$|\langle m\rangle^{}_{\alpha\beta}|$ are different.}.
\begin{itemize}
\item
$\vert{\langle m\rangle^{}_{ee}}\vert$. This is the simplest case, in which
\begin{eqnarray}
\left\vert{\langle m\rangle^{}_{ee}}\right\vert = \left\vert{m^{}_{1} c^{2}_{12} c^{2}_{13}
e^{{\rm i}\rho} + m^{}_{2} s^{2}_{12} c^{2}_{13}e^{{\rm i}\sigma} + m^{}_{3} s^{2}_{13}
}\right\vert
\end{eqnarray}
with $\rho = \xi^{}_{1}+2\delta$ and $\sigma = \xi^{}_{2}+2\delta $. Note that a
different phase assignment has been used for the 3D mapping of
$\vert{\langle m\rangle^{}_{ee}}\vert$ in Refs.~\cite{Xing:2015zha,Xing:2016ymd,Cao:2019hli}. We shall comment on this issue and make a comparison between the two phase
assignments in section 4.2.

\item
$\vert{\langle m\rangle^{}_{\mu\mu}}\vert$. The explicit expression of the modulus of
this diagonal matrix element is given by
\begin{eqnarray}
\vert{\langle m\rangle^{}_{\mu\mu}}\vert = \Biggl\lvert{m^{}_{1} e^{{\rm i}\rho}\left\lvert{\left(s^{}_{12} c^{}_{23} + c^{}_{12} s^{}_{23} s^{}_{13} e^{{\rm i}\delta}\right)^{2}} \right\rvert + m^{}_{2} e^{{\rm i}\sigma} \left\lvert{\left(c^{}_{12} c^{}_{23} - s^{}_{12} s^{}_{23} s^{}_{13} e^{{\rm i}\delta}\right)^{2}} \right\rvert + m^{}_{3} c^{2}_{13} s^2_{23}}\Biggr\rvert \hspace{0.4cm}
\end{eqnarray}
with $\rho = \xi^{}_{1} + \arg\left[\left(s^{}_{12} c^{}_{23} + c^{}_{12} s^{}_{23} s^{}_{13} e^{{\rm i}\delta}\right)^{2}\right]$ and $\sigma = \xi^{}_{2} + \arg\left[\left(c^{}_{12} c^{}_{23} - s^{}_{12} s^{}_{23} s^{}_{13} e^{{\rm i}\delta}\right)^{2}\right]$.

\item
$\vert{\langle m\rangle^{}_{\tau\tau}}\vert$. This diagonal matrix element can be obtained
from $\vert{\langle m\rangle^{}_{\mu\mu}}\vert$ by making the replacements $c^{}_{23} \to
s^{}_{23}$ and $s^{}_{23} \to -c^{}_{23}$. Namely,
\begin{eqnarray}
\vert{\langle m\rangle^{}_{\tau\tau}}\vert = \Biggl\lvert{m^{}_{1} e^{{\rm i}\rho}\left\lvert\left(s^{}_{12} s^{}_{23} - c^{}_{12} c^{}_{23} s^{}_{13} e^{{\rm i}\delta}\right)^{2} \right\rvert + m^{}_{2} e^{{\rm i}\sigma}\left\lvert{\left(c^{}_{12} s^{}_{23} + s^{}_{12} c^{}_{23} s^{}_{13} e^{{\rm i}\delta}\right)^{2}}\right\rvert + m^{}_{3} c^2_{13} c^2_{23}}\Biggr\rvert
\hspace{0.45cm}
\end{eqnarray}
with $\rho = \xi^{}_{1} + \arg\left[\left(s^{}_{12} s^{}_{23} - c^{}_{12} c^{}_{23} s^{}_{13} e^{{\rm i}\delta}\right)^{2}\right]$ and $\sigma = \xi^{}_{2} + \arg\left[\left(c^{}_{12} s^{}_{23} + s^{}_{12} c^{}_{23} s^{}_{13} e^{{\rm i}\delta}\right)^{2}\right]$.

\item
$\vert{\langle m\rangle^{}_{e\mu}}\vert$. This off-diagonal matrix element of $M^{}_\nu$ can be
expressed in the following way:
\begin{eqnarray}
&& \vert{\langle m\rangle^{}_{e\mu}}\vert = \Biggl\lvert
-m^{}_{1} e^{{\rm i}\rho} c^{}_{12} c^{}_{13}\left\lvert{s^{}_{12} c^{}_{23} + c^{}_{12} s^{}_{23} s^{}_{13} e^{{\rm i}\delta}}\right\rvert +
m^{}_{2} e^{{\rm i}\sigma} s^{}_{12} c^{}_{13} \left\lvert{c^{}_{12} c^{}_{23} - s^{}_{12} s^{}_{23} s^{}_{13} e^{{\rm i}\delta}}\right\rvert \hspace{1cm}
\nonumber \\
&& \hspace{1.95cm} + m^{}_{3} c^{}_{13} s^{}_{23} s^{}_{13} \Biggr\rvert
\end{eqnarray}
with $\rho = \xi^{}_{1} + \arg\left(s^{}_{12} c^{}_{23} + c^{}_{12} s^{}_{23} s^{}_{13} e^{{\rm i}\delta}\right) + \delta$ and $\sigma = \xi^{}_{2} + \arg\left(c^{}_{12} c^{}_{23} - s^{}_{12} s^{}_{23} s^{}_{13} e^{{\rm i}\delta}\right) + \delta$.

\item
$\vert{\langle m\rangle^{}_{e\tau}}\vert$. This matrix element can be obtained from
$\vert{\langle m\rangle^{}_{e\mu}}\vert$ by making the replacements $c^{}_{23} \to
s^{}_{23}$ and $s^{}_{23} \to -c^{}_{23}$. Namely,
\begin{eqnarray}
&& \vert{\langle m\rangle^{}_{e\tau}}\vert = \Biggl\lvert +m^{}_{1} e^{{\rm i}\rho} c^{}_{12} c^{}_{13}\left\lvert{s^{}_{12} s^{}_{23} - c^{}_{12} c^{}_{23} s^{}_{13} e^{{\rm i}\delta}}\right\rvert
- m^{}_{2} e^{{\rm i}\sigma} s^{}_{12} c^{}_{13} \left\lvert{c^{}_{12} s^{}_{23} + s^{}_{12} c^{}_{23} s^{}_{13} e^{{\rm i}\delta}}\right\rvert \hspace{1cm}
\nonumber \\
&& \hspace{1.95cm} + m^{}_{3} c^{}_{13} c^{}_{23} s^{}_{13} \Biggr\rvert
\end{eqnarray}
with $\rho = \xi^{}_{1} + \arg\left(s^{}_{12} s^{}_{23} - c^{}_{12} c^{}_{23} s^{}_{13} e^{{\rm i}\delta}\right) + \delta$ and $\sigma = \xi^{}_{2} + \arg\left(c^{}_{12} s^{}_{23} + s^{}_{12} c^{}_{23} s^{}_{13} e^{{\rm i}\delta}\right) + \delta$.

\item
$\vert{\langle m\rangle^{}_{\mu\tau}}\vert$. This off-diagonal matrix element is found
to be unchanged under the interchanges
$c^{}_{23} \to s^{}_{23}$ and $s^{}_{23} \to -c^{}_{23}$. Its explicit expression is
\begin{eqnarray}
&& \vert{\langle m\rangle^{}_{\mu\tau}}\vert = \Biggl\vert - m^{}_{1} e^{{\rm i}\rho}\left\lvert{\left(s^{}_{12} c^{}_{23} + c^{}_{12} s^{}_{23} s^{}_{13} e^{{\rm i}\delta}\right)\left(s^{}_{12} s^{}_{23} - c^{}_{12} c^{}_{23} s^{}_{13} e^{{\rm i}\delta}\right)}\right\rvert
\nonumber \\
&& \hspace{1.95cm}
- m^{}_{2} e^{{\rm i}\sigma}\left\lvert{\left(c^{}_{12} c^{}_{23} - s^{}_{12} s^{}_{23} s^{}_{13} e^{{\rm i}\delta}\right)\left(c^{}_{12} s^{}_{23} + s^{}_{12} c^{}_{23} s^{}_{13} e^{{\rm i}\delta}\right)}\right\rvert + m^{}_{3} c^{2}_{13} c^{}_{23} s^{}_{23} \Biggr\vert \hspace{1.5cm}
\end{eqnarray}
with the two effective phases
$\rho = \xi^{}_{1} + \arg\left[\left(s^{}_{12} c^{}_{23} + c^{}_{12} s^{}_{23} s^{}_{13} e^{{\rm i}\delta}\right)\left(s^{}_{12} s^{}_{23} - c^{}_{12} c^{}_{23} s^{}_{13} e^{{\rm i}\delta}\right)\right]$ and  $\sigma = \xi^{}_{2} + \arg\left[\left(c^{}_{12} c^{}_{23} - s^{}_{12} s^{}_{23} s^{}_{13} e^{{\rm i}\delta}\right)\left(c^{}_{12} s^{}_{23} + s^{}_{12} c^{}_{23} s^{}_{13} e^{{\rm i}\delta}\right)\right]$.
\end{itemize}
One can see that the effective Majorana CP phases $\rho$ and $\sigma$ are linearly related to
the original Majorana phases $\xi^{}_1$ and $\xi^{}_2$ of the PMNS matrix $U$,
and the phase differences $\rho-\xi_{1}$ and $\sigma-\xi_{2}$ depend upon the
Dirac phase $\delta$ and the relevant neutrino mixing angles. Given the fact that
the values of $\theta^{}_{12}$, $\theta^{}_{13}$ and $\theta^{}_{23}$ have all been
determined to a good degree of accuracy, it is easy to show the dependence of
$\rho-\xi_{1}$ and $\sigma-\xi_{2}$ on $\delta$ in a numerical way.

\begin{figure}[htbp]
\centering
{
\includegraphics[scale=1.38]{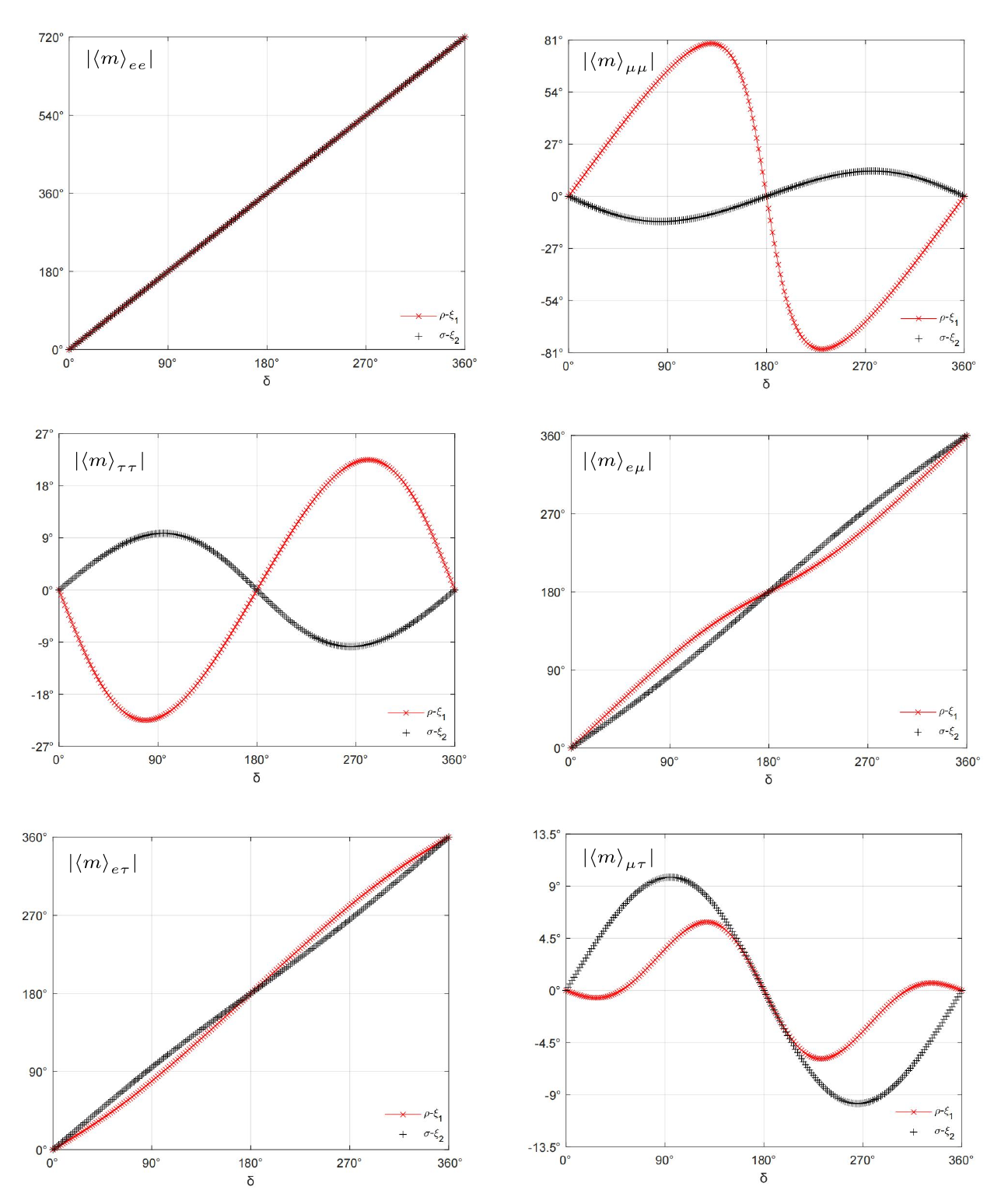}
}
\vspace{-0.8cm}
\caption{The variations of $\rho-\xi^{}_{1}$ and $\sigma-\xi^{}_{2}$ with respect to
$\delta$ for each of $|\langle m\rangle^{}_{\alpha\beta}|$ in the NMO case, where
the best-fit values of three neutrino mixing angles have been input.}
\label{FIG1}
\end{figure}

Let us illustrate the changes of $\rho-\xi^{}_{1}$ and $\sigma-\xi^{}_{2}$ against
$\delta$ by inputting the best-fit values of the two independent neutrino
mass-squared differences and three flavor mixing angles in either the NMO case or the
IMO case~\cite{Esteban:2020cvm}
\footnote{Although the best-fit value of $\delta$ has also been obtained from a global
analysis of current neutrino oscillation data, namely $\delta = 195^{\circ}$ (NMO) or $286^{\circ}$ (IMO)~\cite{Esteban:2020cvm}, it remains quite uncertain due to the poor
statistical significance.}:
$\Delta m^{2}_{21} = 7.42 \times 10^{-5}$ eV, $\Delta m^{2}_{31} = 2.514 \times 10^{-3}$
eV (NMO) or $\Delta m^{2}_{32} = -2.497 \times 10^{-3}$ eV (IMO),
$\theta^{}_{12} = 33.44^{\circ}$ (NMO) or $33.45^{\circ}$ (IMO),
$\theta^{}_{13} = 8.57^{\circ}$ (NMO) or $8.61^{\circ}$ (IMO), and
$\theta^{}_{23} = 49.0^{\circ}$ (NMO) or $49.3^{\circ}$ (IMO). Fig.~\ref{FIG1}
shows the numerical results for the NMO case, and those for the IMO case are
found to be rather similar.
It is obvious that the phase differences $\rho-\xi^{}_{1}$ and $\sigma-\xi^{}_{2}$
associated with $\vert{\langle m\rangle^{}_{ee}}\vert$ depend linearly on $\delta$;
those associated with $\vert{\langle m\rangle^{}_{e\mu}}\vert$ and
$\vert{\langle m\rangle^{}_{e\tau}}\vert$ vary almost linearly with $\delta$
as a result of the smallness of $\theta^{}_{13}$; and those associated with
the other three effective Majorana neutrino masses oscillate periodically with $\delta$.

\subsection{The upper and lower bounds of $|\langle m\rangle^{}_{\alpha\beta}|$}

Now we proceed to analytically determine the extremum of each
$|\langle m\rangle^{}_{\alpha\beta}|$ against the effective Majorana phase $\sigma$,
which is related to $m^{}_2$ and thus less sensitive to the mass ordering of three
neutrinos. To this end, we express $|\langle m\rangle^{}_{\alpha\beta}|$ in a
generic way as follows:
\begin{eqnarray}
\vert{\langle m\rangle^{}_{\alpha\beta}}\vert = \left| r^{}_{1} e^{{\rm i} \rho}
+ r^{}_{2} e^{{\rm i}\sigma} + r^{}_{3} \right| \; ,
\end{eqnarray}
where $r^{}_{i} \propto m^{}_i$ (for $i=1,2,3$) are real. Allowing Eq.~(12) to take
its extremum with respect to $\sigma$,
\begin{eqnarray}
\frac{\partial \vert{\langle m\rangle^{}_{\alpha\beta}}\vert}
{\partial \sigma} = 0 \;\; \Longrightarrow \;\;
\frac{\displaystyle r^{}_{1} r^{}_{2} \sin\left(\rho - \sigma\right) - r^{}_{2} r^{}_{3}
\sin \sigma}{\displaystyle \sqrt{r^{2}_{1} + r^{2}_{2} + r^{2}_{3} + 2 r^{}_{1} r^{}_{2} \cos\left(\rho-\sigma\right) + 2 r^{}_{1} r^{}_{3} \cos \rho + 2 r^{}_{2} r^{}_{3}
\cos\sigma}} = 0 \; ,
\end{eqnarray}
we obtain the condition
\begin{eqnarray}
\tan\sigma = \frac{r^{}_{1}\sin\rho}{r^{}_{1}\cos{\rho} + r^{}_{3}} \; ,
\end{eqnarray}
which permits us to either maximize or minimize the magnitude of
$\vert{\langle m\rangle^{}_{\alpha\beta}}\vert$
for the given values of $\rho$ and $m^{}_1$ (or $m^{}_3$).
Substituting Eq.~(14) into Eq.~(12), we immediately arrive at the upper (``U") and
lower (``L") bounds of $\vert{\langle m\rangle^{}_{\alpha\beta}}\vert$:
\begin{eqnarray}
&& \vert{\langle m\rangle^{}_{\alpha\beta}}\vert^{}_{\rm U} = \Biggl\lvert{\left| r^{}_{1}
e^{{\rm i}\rho} + r^{}_{3} \right| + r^{}_{2}}\Biggr\rvert \; ,
\nonumber \\
&& \vert{\langle m\rangle^{}_{\alpha\beta}}\vert^{}_{\rm L} = \Biggl\lvert{\left| r^{}_{1}
e^{{\rm i}\rho} + r^{}_{3} \right| - r^{}_{2}}\Biggr\rvert \; . \hspace{1.5cm}
\end{eqnarray}
These two bounds will help us a lot in determining the bulk of the parameter space of
$\vert{\langle m\rangle^{}_{\alpha\beta}}\vert$ for arbitrary values of $\rho$ and
reasonable values of $m^{}_1$ (or $m^{}_3$). In this work we focus on
$m^{}_i \lesssim 0.1~{\rm eV}$, as conservatively constrained by current observational data~\cite{Zyla:2020zbs}.

To be more specific, we write out the upper and lower bounds for each of the six effective
Majorana neutrino masses as follows.
\begin{itemize}
\item
$\vert{\langle m\rangle^{}_{ee}}\vert$.
Defining $m^{ee}_{1} = m^{}_{1} c^{2}_{12} c^{2}_{13}$, $m^{ee}_{2} = m^{}_{2} s^{2}_{12} c^{2}_{13}$ and $m^{ee}_{3} = m^{}_{3} s^{2}_{13}$, we have
\begin{eqnarray}
\vert{\langle m\rangle^{}_{ee}}\vert^{}_{\rm U,L} = \Biggl\lvert
\left| m^{ee}_{1} e^{{\rm i}\rho} + m^{ee}_{3}\right| \pm m^{ee}_{2}\Biggr\rvert \; .
\hspace{1cm}
\end{eqnarray}

\item
$\vert{\langle m\rangle_{\mu\mu}}\vert$.
Defining $m^{\mu\mu}_{1} = m^{}_{1} \left|\left(s^{}_{12} c^{}_{23} + c^{}_{12} s^{}_{23} s^{}_{13} e^{{\rm i}\delta}\right)^{2}\right|$, $m^{\mu\mu}_{2} = m^{}_{2}\left|\left(c^{}_{12} c^{}_{23} - s^{}_{12} s^{}_{23} s^{}_{13} e^{{\rm i}\delta}\right)^{2}\right|$ and
$m^{\mu\mu}_{3} = m^{}_{3} c^{2}_{13} s^{2}_{23}$, we obtain
\begin{eqnarray}
\vert{\langle m\rangle^{}_{\mu\mu}}\vert^{}_{\rm U,L} =
\Biggl\lvert \left| m^{\mu\mu}_{1} e^{{\rm i}\rho} + m^{\mu\mu}_{3}\right| \pm m^{\mu\mu}_{2}\Biggr\rvert \; . \hspace{1cm}
\end{eqnarray}

\item
$\vert{\langle m\rangle^{}_{\tau\tau}}\vert$.
Defining $m^{\tau\tau}_{1} = m^{}_{1} \left|\left(s^{}_{12} s^{}_{23} - c^{}_{12} c^{}_{23} s^{}_{13} e^{{\rm i}\delta}\right)^{2}\right|$, $m^{\tau\tau}_{2} = m^{}_{2}\left|\left(c^{}_{12} s^{}_{23} + s^{}_{12} c^{}_{23} s^{}_{13} e^{{\rm i}\delta}\right)^{2}\right|$ and
$m^{\tau\tau}_{3} = m^{}_{3}c^{2}_{13} c^{2}_{23}$, we arrive at
\begin{eqnarray}
\vert{\langle m\rangle^{}_{\tau\tau}}\vert^{}_{\rm U,L} = \Biggl\lvert\left| m^{\tau\tau}_{1} e^{{\rm i}\rho} + m^{\tau\tau}_{3}\right| \pm m^{\tau\tau}_{2}\Biggr\rvert \; .
\hspace{1cm}
\end{eqnarray}

\item
$\vert{\langle m\rangle^{}_{e\mu}}\vert$.
Defining $m^{e\mu}_{1} = m^{}_{1} c^{}_{12} c^{}_{13} \lvert{s^{}_{12} c^{}_{23} + c^{}_{12} s^{}_{23} s^{}_{13} e^{{\rm i}\delta}}\rvert$,
$m^{e\mu}_{2} = m^{}_{2} s^{}_{12} c^{}_{13} \lvert{c^{}_{12} c^{}_{23} - s^{}_{12} s^{}_{23} s^{}_{13} e^{{\rm i}\delta}}\rvert$ and
$m^{e\mu}_{3} = m^{}_{3} c^{}_{13} s^{}_{23} s^{}_{13}$, we have
\begin{eqnarray}
\vert{\langle m\rangle^{}_{e\mu}}\vert^{}_{\rm U,L} = \Biggl\lvert\left| - m^{e\mu}_{1} e^{{\rm i}\rho} + m^{e\mu}_{3}\right| \pm m^{e\mu}_{2}\Biggr\rvert \; . \hspace{1cm}
\end{eqnarray}

\item
$|\left \langle m\rangle^{}_{e\tau}\right|$.
Defining $m^{e\tau}_{1} = m^{}_{1} c^{}_{12} c^{}_{13}\left| s^{}_{12} s^{}_{23} - c^{}_{12} c^{}_{23} s^{}_{13} e^{{\rm i}\delta}\right|$, $m^{e\tau}_{2} = m^{}_{2} s^{}_{12} c^{}_{13}\left| c^{}_{12} s^{}_{23} + s^{}_{12} c^{}_{23} s^{}_{13} e^{{\rm i}\delta}\right|$
and $m^{e\tau}_{3} = m^{}_{3} c^{}_{13} c^{}_{23} s^{}_{13}$, we obtain
\begin{eqnarray}
\vert{\langle m\rangle^{}_{e\tau}}\vert^{}_{\rm U,L} = \Biggl\lvert\left| m^{e\tau}_{1} e^{{\rm i}\rho} + m^{e\tau}_{3}\right| \pm m^{e\tau}_{2}\Biggr\rvert \; . \hspace{1cm}
\end{eqnarray}

\item
$\vert{\langle m\rangle^{}_{\mu\tau}}\vert$.
Let us define $m^{\mu\tau}_{1} = m^{}_{1}\left|\left(s^{}_{12}c_{23} + c^{}_{12} s^{}_{23} s^{}_{13}
e^{{\rm i}\delta}\right)\left(s^{}_{12} s^{}_{23} - c^{}_{12} c^{}_{23} s^{}_{13} e^{{\rm i}\delta}\right)\right|$ together with
$m^{\mu\tau}_{2} = m^{}_{2}\left|\left(c^{}_{12} c^{}_{23} - s^{}_{12} s^{}_{23} s^{}_{13} e^{{\rm i}\delta}\right)\left(c^{}_{12} s^{}_{23} + s^{}_{12} c^{}_{23} s^{}_{13} e^{{\rm i}\delta}\right)\right|$ and
$m^{\mu\tau}_{3} = m^{}_{3} c^{2}_{13} c^{}_{23} s^{}_{23}$. Then
\begin{eqnarray}
\vert{\langle m\rangle^{}_{\mu\tau}}\vert^{}_{\rm U,L} = \Biggl\lvert\left|  -m^{\mu\tau}_{1} e^{{\rm i}\rho} + m^{\mu\tau}_{3}\right| \pm m^{\mu\tau}_{2}\Biggr\rvert \; . \hspace{1cm}
\end{eqnarray}
\end{itemize}
These analytical results will be used in our numerical mapping of the 3D profiles of
$\vert{\langle m\rangle^{}_{\alpha\beta}}\vert$.

\section{Numerical 3D mapping of $\vert{\langle m\rangle^{}_{\alpha\beta}}\vert$}

\subsection{Normal mass ordering}

\begin{figure}[htbp]
\centering
{
\includegraphics[scale=1.34]{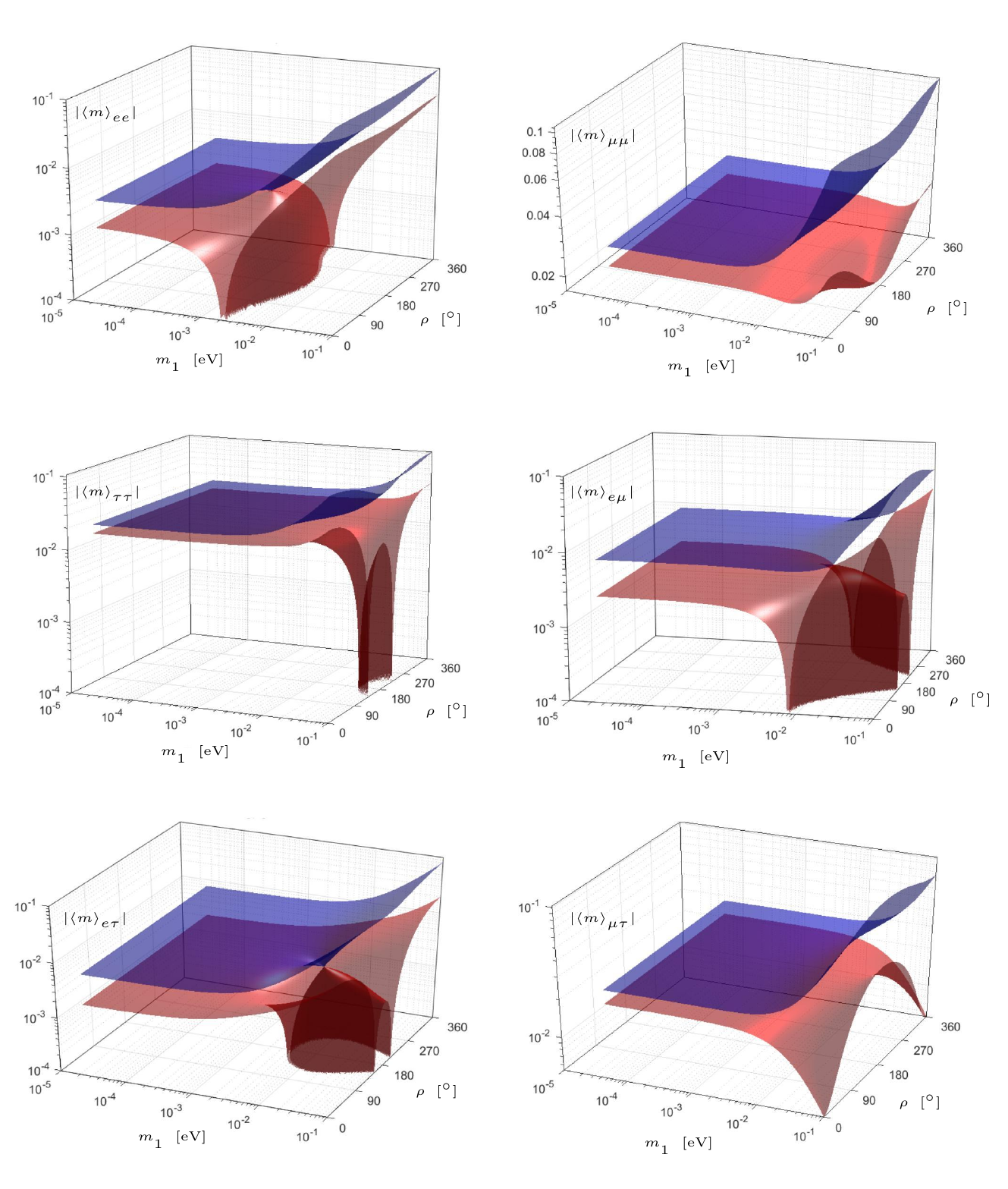}
}
\vspace{-1cm}
\caption{The 3D profile of each of the six effective Majorana neutrino masses
$|\langle m\rangle^{}_{\alpha\beta}|$ (for $\alpha, \beta = e, \mu, \tau$):
its upper (blue) and lower (red) bounds are functions of $m^{}_1$ and $\rho$
in the NMO case. Note that the two touching points between the upper and lower
layers of $|\langle m\rangle^{}_{e\mu}|$ are equivalent in physics, because they
correspond to $\rho = 0$ and $2\pi$.}
\label{FIG2}
\end{figure}

In the NMO case (i.e., $m^{}_1 < m^{}_2 < m^{}_3$) we choose the smallest neutrino mass
$m^{}_1$ as a free parameter. The other two neutrino masses can then be expressed as
$m^{}_{2} = \sqrt{{m^2_{1}} + \Delta m^{2}_{21}}$ and
$m^{}_{3} = \sqrt{{m^2_{1}} + \Delta m^{2}_{31}}$. With the help of Eqs.~(16)---(21) and the
best-fit values of $\Delta m^2_{21}$, $\Delta m^2_{31}$, $\theta^{}_{12}$, $\theta^{}_{13}$,
$\theta^{}_{23}$ and $\delta$ listed in section 2~\cite{Esteban:2020cvm}, we
plot the minimum and maximum of each of $|\langle m\rangle^{}_{\alpha\beta}|$
with respect to the unknown phase parameter $\sigma$ by requiring the other unknown
phase parameter $\rho$ and the unknown neutrino mass $m^{}_1$ to vary in their
respectively allowed regions. Our numerical results for the 3D profiles of
$|\langle m\rangle^{}_{\alpha\beta}|$ are shown in Fig.~\ref{FIG2}. Some explicit discussions
are in order.

\subsubsection{The bounds of $\vert{\langle m\rangle^{}_{ee}}\vert$}

As shown in Fig.~\ref{FIG2}, there is a unique touching point between the upper
and lower layers of $\vert{\langle m\rangle^{}_{ee}}\vert$.
The location of this interesting point can be analytically fixed from Eq.~(16)
by simply setting $\vert{\langle m\rangle^{}_{ee}}\vert^{}_{\rm U} =
\vert{\langle m\rangle^{}_{ee}}\vert^{}_{\rm L} \equiv
\vert{\langle m\rangle^{}_{ee}}\vert^{}_*$. The latter implies
\begin{eqnarray}
m^{}_{1} c^{2}_{12} c^{2}_{13} e^{{\rm i}\rho} + m^{}_{3} s^{2}_{13} = 0 \; ,
\end{eqnarray}
from which we immediately arrive at $\rho = \pi$,
\begin{eqnarray}
m^{}_1 = \sqrt{\frac{\Delta m^2_{31} t^4_{13}}{c^4_{12} - t^4_{13}}}
\simeq \sqrt{\Delta m^2_{31}}  \hspace{0.1cm} \frac{t^2_{13}}{c^2_{12}} \; ,
\end{eqnarray}
and
\begin{eqnarray}
\vert{\langle m\rangle^{}_{ee}}\vert^{}_* = m^{}_2 s^2_{12} c^2_{13}
\simeq \sqrt{\Delta m^2_{21} s^4_{12} c^4_{13} + \Delta m^2_{31}
t^4_{12} s^4_{13}} \; ,
\end{eqnarray}
where $t^{}_{ij} \equiv \tan\theta^{}_{ij}$ (for $ij = 12, 13, 23$).
Inputting the best-fit values of $\theta^{}_{12}$, $\theta^{}_{13}$,
$\Delta m^2_{21}$ and $\Delta m^2_{31}$, we obtain
$\left(\rho, m^{}_{1}\right )^{}_* \simeq \left(\pi, 1.68~{\rm meV}\right)$ and
$\vert{\langle m\rangle^{}_{ee}}\vert^{}_* \simeq 2.61~{\rm meV}$ for the touching point.
In this special case the original Majorana phase $\xi^{}_1$ is related to
$\delta$ through $\xi^{}_{1} = \pi - 2\delta$.
\begin{figure}
\centering
\includegraphics[scale=0.44]{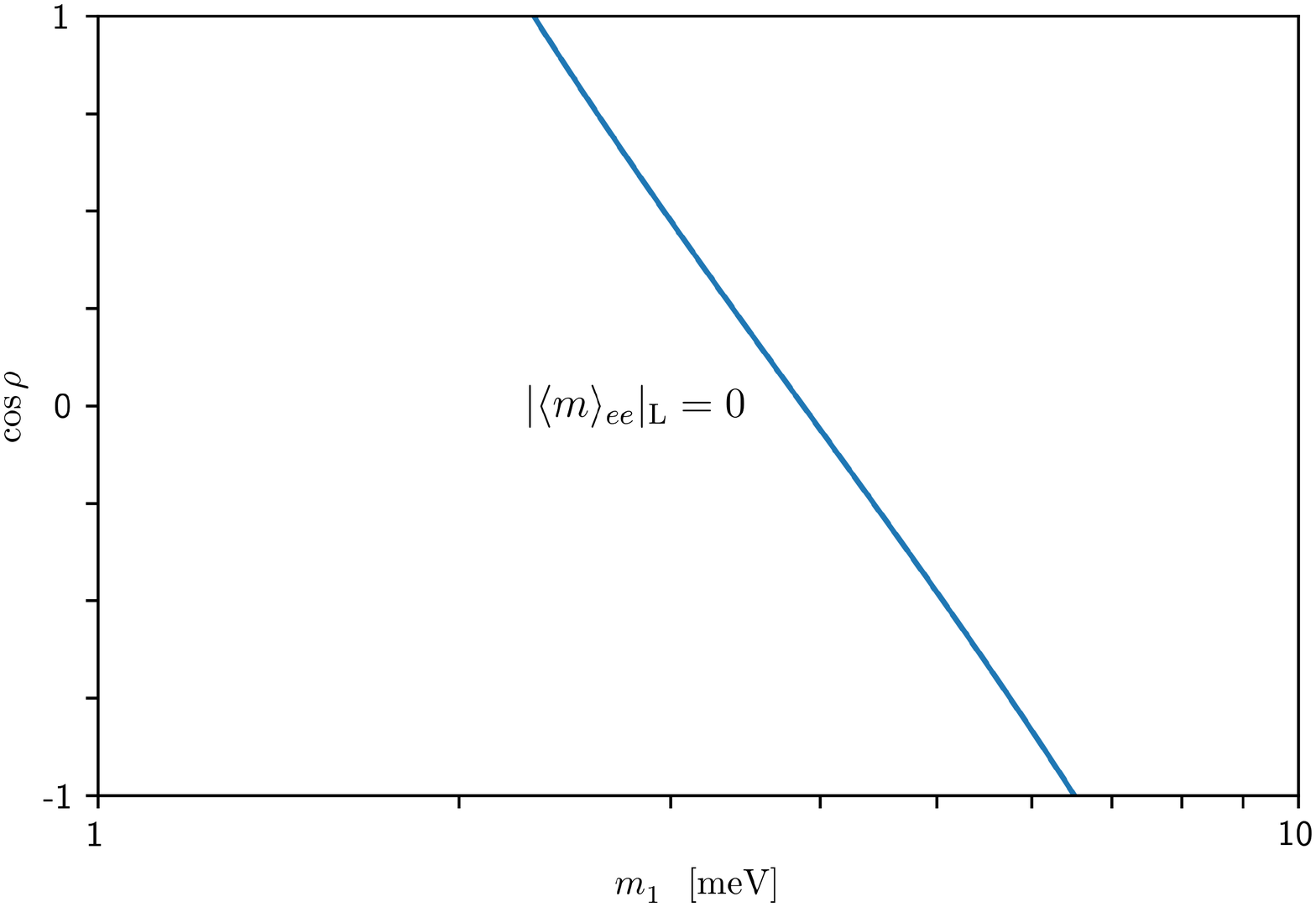}
\vspace{-0.9cm}
\caption{The correlation between $\cos \rho$ and $m^{}_{1}$ constrained by
$\vert{\langle m\rangle^{}_{ee}}\vert^{}_{\rm L}=0$ in the NMO case.}
\label{FIG3}
\end{figure}

Note that $\vert{\langle m\rangle^{}_{ee}}\vert^{}_{\rm L} = 0$ will hold if the
condition
\begin{eqnarray}
\cos\rho = \frac{\displaystyle\left(m^2_2 s^4_{12} - m^2_1 c^4_{12}\right)c^4_{13}
- m^2_3 s^4_{13}}{\displaystyle 2 m^{}_1 m^{}_3 c^2_{12} c^2_{13} s^2_{13}} \;
\end{eqnarray}
is satisfied, as one can easily see from Eq.~(16). This condition allows
us to illustrate the specific correlation between $\cos\rho$ and
$m^{}_1$ in Fig.~\ref{FIG3}. It is clear that $\vert{\langle m\rangle^{}_{ee}}\vert=0$
puts a strict constraint on the range of $m^{}_1$, which approximately varies between
$2 ~{\rm meV}$ and $7 ~{\rm meV}$.

\subsubsection{The bounds of $\vert{\langle m\rangle^{}_{\mu\mu}}\vert$}

As is shown in Fig.~\ref{FIG2}, the upper and lower layers of
$\vert{\langle m\rangle^{}_{\mu\mu}}\vert$ are highly plain or flat for
$m^{}_1 \lesssim 0.05~{\rm eV}$ and arbitrary values of $\rho$, and
they have no intersecting point or area. For $m^{}_1 \lesssim 0.1~{\rm eV}$,
the possibility of $\vert{\langle m\rangle^{}_{\mu\mu}}\vert=0$ has been
ruled out.

\subsubsection{The bounds of $\vert{\langle m\rangle^{}_{\tau\tau}}\vert$}

\begin{figure}
\centering
\includegraphics[scale=0.44]{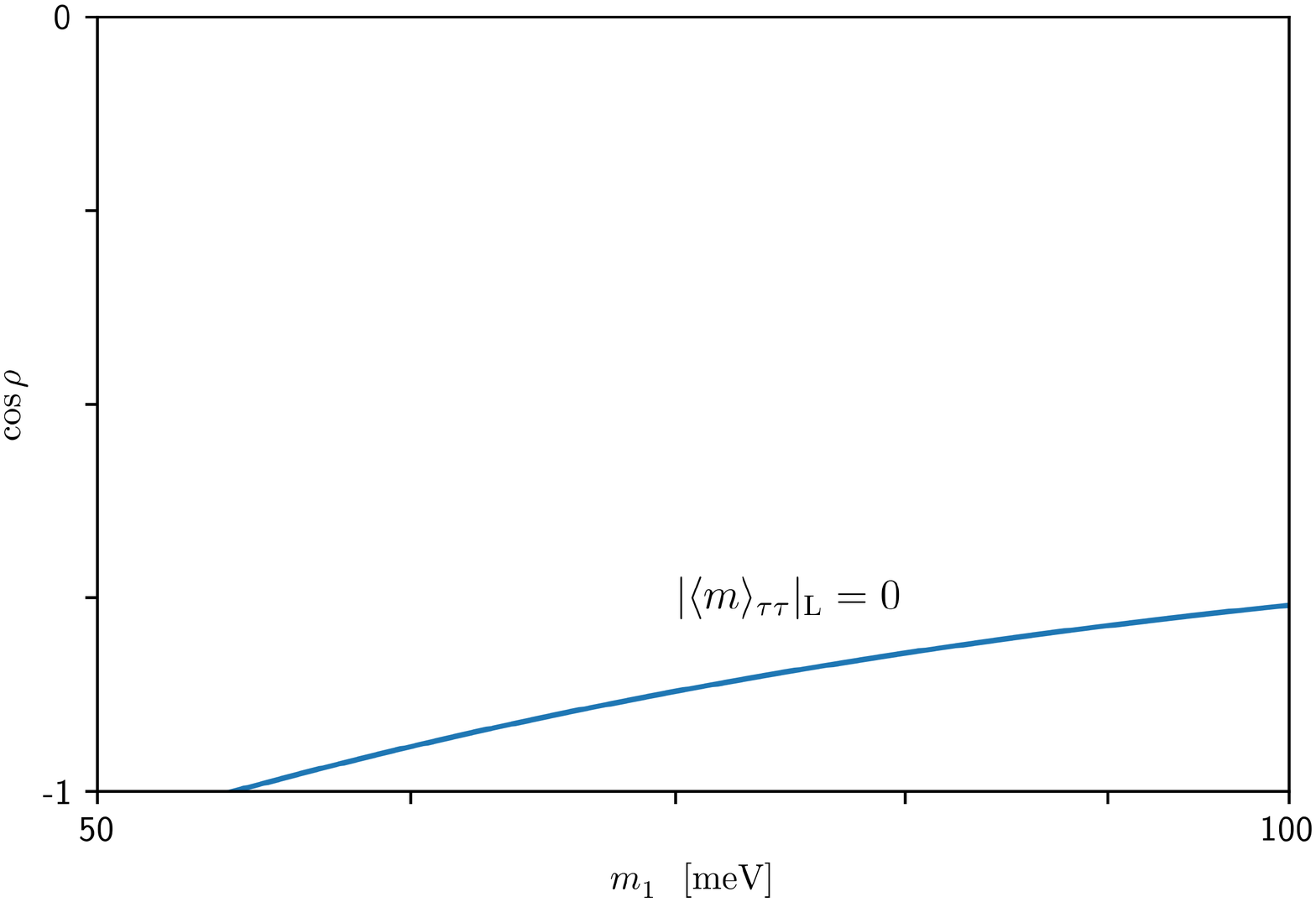}
\vspace{-0.9cm}
\caption{The correlation between $\cos \rho$ and $m^{}_{1}$ constrained by
$\vert{\langle m\rangle^{}_{\tau\tau}}\vert^{}_{\rm L}=0$ in the NMO case.}
\label{FIG4}
\end{figure}

Fig.~\ref{FIG2} tells us that $\vert{\langle m\rangle^{}_{\tau\tau}}\vert^{}_{\rm L}$
is possible to vanish when $m^{}_1$ and $\rho$ approach $0.1~{\rm eV}$ and $\pi$,
respectively. Taking $\vert{\langle m\rangle^{}_{\tau\tau}}\vert^{}_{\rm L} = 0$,
we immediately arrive at
\begin{eqnarray}
\cos \rho=\frac {m^2_2 \left|\left(c^{}_{12}s^{}_{23}+s^{}_{12}c^{}_{23}s^{}_{13}e^{\rm i\delta}\right)^2\right|^2-m^2_{1}\left|\left(s^{}_{12}s^{}_{23}-c^{}_{12}c^{}_{23}
s^{}_{13}e^{\rm i\delta}\right)^2\right|^2 - m^2_{3}c^4_{13}c^4_{23}}{2m^{}_{1}m^{}_{3}c^2_{13}c^2_{23}
\left|\left(s^{}_{12}s^{}_{23}-c^{}_{12}c^{}_{23}s^{}_{13}e^{\rm i\delta}\right)^2\right|} \;
\end{eqnarray}
from Eq.~(18). In this case the correlation between $\cos\rho$ and $m^{}_1$ is illustrated
in Fig.~\ref{FIG4}. It is obvious that
$\vert{\langle m\rangle^{}_{\tau\tau}}\vert^{}_{\rm L} = 0$ sets a lower limit for
$m^{}_{1}$ (roughly about $0.05 ~{\rm eV}$) and an upper limit for
$\cos\rho$ (roughly around $-0.75$). Of course, these numerical observations depend
closely on the best-fit inputs of current neutrino oscillation parameters, but they
may at least give us a ball-park feeling of the interesting correlative behaviors
of $m^{}_1$ and $\cos\rho$ under the condition of
$\vert{\langle m\rangle^{}_{\tau\tau}}\vert^{}_{\rm L} = 0$ in the NMO case.

\subsubsection{The bounds of $\vert{\langle m\rangle^{}_{e\mu}}\vert$}

Two features of the profile of $\vert{\langle m\rangle^{}_{e\mu}}\vert$ in Fig.~\ref{FIG2} are worthy of remarking. Similar to $\vert{\langle m\rangle^{}_{\tau\tau}}\vert^{}_{\rm L}$,
the lower bound of $\vert{\langle m\rangle^{}_{e\mu}}\vert$ is possible to vanish
when $m^{}_{1}$ is around $0.01~{\rm eV}$ or larger. On the other hand,
the upper and lower limits of $\vert{\langle m\rangle^{}_{e\mu}}\vert$ have a unique
touching point fixed by the equality $|\langle m\rangle^{}_{e\mu}|^{}_{\rm U} =
|\langle m\rangle^{}_{e\mu}|^{}_{\rm L} \equiv |\langle m\rangle^{}_{e\mu}|^{}_*$ or
equivalently the condition
\begin{eqnarray}
-m^{}_{1}e^{\rm i\rho}c_{12}\left|s^{}_{12}c^{}_{23}
+ c^{}_{12}s^{}_{23}s^{}_{13}e^{\rm i\delta}\right|
+ m^{}_{3}s^{}_{13}s^{}_{23} = 0 \; .
\end{eqnarray}
So it is straightforward for us to obtain $\rho=0$ (or $2\pi$)
from Eq.~(27), together with
\begin{eqnarray}
m^{}_1 = \sqrt{\frac{\Delta m^2_{31} s^2_{13}s^2_{23}}{\displaystyle
c^2_{12}\lvert{s^{}_{12}c^{}_{23}+c^{}_{12}s^{}_{23}s^{}_{13}
e^{\rm i\delta}}\rvert^2 - s^2_{13}s^2_{23}}} \; ,
\end{eqnarray}
and
\begin{eqnarray}
\vert{\langle m\rangle^{}_{e\mu}}\vert^{}_* = m^{}_2 s^{}_{12} c^{}_{13}
\left|c^{}_{12} c^{}_{23} - s^{}_{12} s^{}_{23} s^{}_{13} e^{{\rm i}\delta}\right|
= \sqrt{m^2_1+\Delta m^2_{21}} \hspace{0.05cm}
s^{}_{12} c^{}_{13} \left|c^{}_{12} c^{}_{23} - s^{}_{12} s^{}_{23} s^{}_{13} e^{{\rm i}\delta}\right| \; .
\end{eqnarray}
Numerically, we find $(\rho, m^{}_{1}) \simeq (0, 0.0289~{\rm eV})$ and
$\vert{\langle m\rangle^{}_{e\mu}}\vert^{}_* \simeq 9.88 ~{\rm meV}$ for
the touching point, where the original Majorana phase $\xi^{}_1$ reads as
$\xi^{}_{1} = -\arg\left(s^{}_{12}c^{}_{23} + c^{}_{12}s^{}_{23}s^{}_{13}
e^{\rm i\delta}\right) - \delta$.

\begin{figure}
\centering
\includegraphics[scale=0.44]{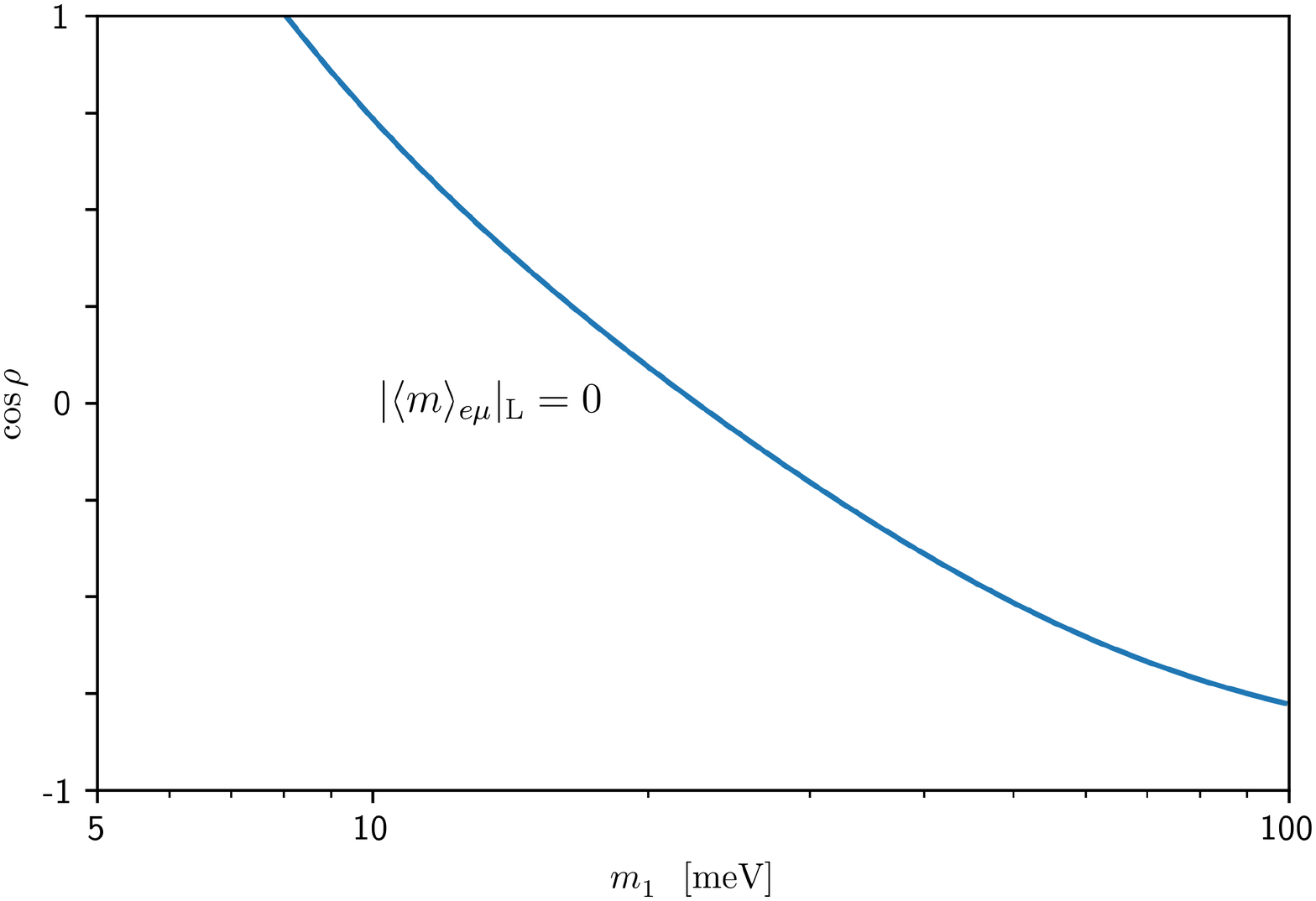}
\vspace{-0.9cm}
\caption{The correlation between $\cos \rho$ and $m^{}_{1}$ constrained by
$\vert{\langle m\rangle^{}_{e\mu}}\vert^{}_{\rm L}=0$ in the NMO case.}
\label{FIG5}
\end{figure}

Moreover, $\vert{\langle m\rangle^{}_{e\mu}}\vert^{}_{\rm L} = 0$ can be obtained
if the condition
\begin{eqnarray}
\cos \rho = \frac {m^2_{3}s^2_{13}s^2_{23}+m^2_{1}c^2_{12}\vert s^{}_{12}c^{}_{23}+c^{}_{12}s^{}_{23}s^{}_{13} e^{\rm i\delta}\vert^2-
m^2_{2}s^2_{12}\vert c^{}_{12}c^{}_{23}-s^{}_{12}s^{}_{23}s^{}_{13}
e^{\rm i\delta}\vert^2}{2m^{}_1m^{}_3c^{}_{12}s^{}_{13}s_{23}\vert s^{}_{12}c^{}_{23}+c^{}_{12}s^{}_{23}s^{}_{13}
e^{\rm i\delta}\vert}
\end{eqnarray}
is satisfied. A numerical illustration of this condition is presented in
Fig.~\ref{FIG5}, from which one may also read out the lower bounds of
$m^{}_1$ and $\cos\rho$ for $\vert{\langle m\rangle^{}_{e\mu}}\vert^{}_{\rm L} = 0$.

\subsubsection{The bounds of $\vert{\langle m\rangle^{}_{e\tau}}\vert$}

Fig.~\ref{FIG2} shows that the 3D profile of $\vert{\langle m\rangle^{}_{e\tau}}\vert$
has an interesting  bullet-like structure, whose tip is just the touching point
between the layers of $\vert{\langle m\rangle^{}_{e\tau}}\vert^{}_{\rm U}$ and
$\vert{\langle m\rangle^{}_{e\tau}}\vert^{}_{\rm L}$. In addition, one can clearly see that
$\vert{\langle m\rangle^{}_{e\tau}}\vert$ is possible to vanish when $\rho$ and
$m^{}_1$ take some appropriate values. The equality
$\vert{\langle m\rangle^{}_{e\tau}}\vert^{}_{\rm U} =
\vert{\langle m\rangle^{}_{e\tau}}\vert^{}_{\rm L} \equiv
\vert{\langle m\rangle^{}_{e\tau}}\vert^{}_*$ requires
\begin{eqnarray}
m^{}_{1}e^{\rm i\rho}c^{}_{12}\lvert s^{}_{12}s^{}_{23}-c^{}_{12}c^{}_{23}s^{}_{13}e^{\rm i\delta}\rvert+m^{}_{3}s^{}_{13}c^{}_{23} = 0 \;
\end{eqnarray}
to hold. As a result, the touching point is fixed by $\rho=\pi$ and
\begin{eqnarray}
m^{}_1 = \sqrt{\frac{\Delta m^2_{31} s^2_{13}c^2_{23}}{\displaystyle c^2_{12}\lvert s^{}_{12}s^{}_{23}-c^{}_{12}c^{}_{23}s^{}_{13}e^{\rm i\delta}\rvert^2
- s^2_{13}c^2_{23}}} \; ,
\end{eqnarray}
and thus
\begin{eqnarray}
\vert{\langle m\rangle^{}_{e\tau}}\vert^{}_* = m^{}_2 s^{}_{12} c^{}_{13} \lvert{c^{}_{12} s^{}_{23} + s^{}_{12} c^{}_{23} s^{}_{13} e^{\rm i\delta}}\rvert
= \sqrt{m^2_{1}+\Delta m^2_{21}} \hspace{0.05cm}
s^{}_{12} c^{}_{13} \lvert{c^{}_{12} s^{}_{23} + s^{}_{12} c^{}_{23} s^{}_{13}
e^{{\rm i}\delta}}\rvert \; .
\end{eqnarray}
Comparing Eqs.~(32) and (33) with Eqs.~(28) and (29), we find that the former can be
easily acquired from the latter by making the replacements
$c^{}_{23} \to s^{}_{23}$ and $s^{}_{23} \to -c^{}_{23}$. This observation implies
a kind of (approximate) $\mu$-$\tau$ interchange symmetry between
$\vert{\langle m\rangle^{}_{e\mu}}\vert$ and $\vert{\langle m\rangle^{}_{e\tau}}\vert$,
given the fact of $\theta^{}_{23} \simeq \pi/4$ as indicated by current neutrino
oscillation data \cite{Xing:2015fdg}. Numerically, we have
$(\rho, m^{}_{1}) \simeq (\pi, 0.0124 ~{\rm eV})$ and $\vert{\langle m\rangle^{}_{e\tau}}\vert^{}_* \simeq 4.69{~\rm meV}$ for the touching point, where
the original Majorana phase $\xi^{}_1$ is related to $\delta$ through the relation
$\xi^{}_{1} = \pi - \arg\left(s_{12}s_{23}-c_{12}c_{23}s_{13}e^{\rm i\delta}\right)
-\delta$.
\begin{figure}
\centering
\includegraphics[scale=0.44]{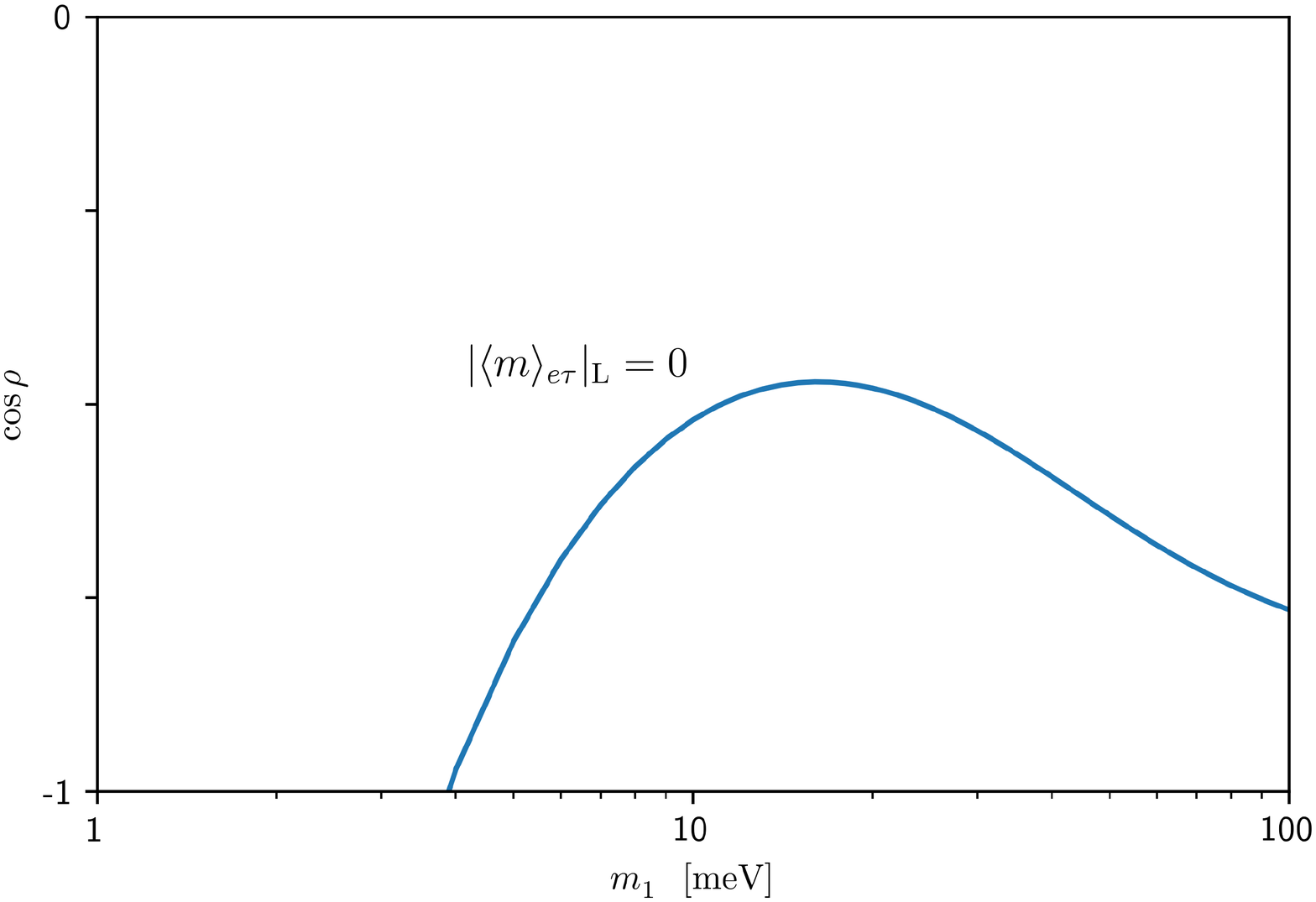}
\vspace{-0.9cm}
\caption{The correlation between $\cos \rho$ and $m^{}_{1}$ constrained by
$\vert{\langle m\rangle^{}_{e\tau}}\vert^{}_{\rm L}=0$ in the NMO case.}
\label{FIG6}
\end{figure}

On the other hand, $\vert{\langle m\rangle^{}_{e\tau}}\vert^{}_{\rm L}=0$ will hold
under the condition of
\begin{eqnarray}
\cos \rho = \frac{\displaystyle m^2_{2}s^2_{12}\vert c^{}_{12}s^{}_{23} +s^{}_{12}c^{}_{23}s^{}_{13}e^{\rm i\delta}\vert^2-m^2_{3}c^2_{23}s^2_{13}
-m^2_{1}c^2_{12}\vert s^{}_{12}s^{}_{23}-c^{}_{12}c^{}_{23}s^{}_{13}e^{\rm i\delta}\vert^2} {\displaystyle 2m^{}_1m^{}_3c^{}_{12}s^{}_{13}c_{23}\vert s^{}_{12}s^{}_{23}
-c^{}_{12}c^{}_{23}s^{}_{13}e^{\rm i\delta}\vert} \; .
\end{eqnarray}
A numerical illustration of this condition is shown in Fig.~\ref{FIG6}, in which
one can see a lower bound $m^{}_1 \gtrsim 4~{\rm meV}$ as required by
$\vert{\langle m\rangle^{}_{e\tau}}\vert^{}_{\rm L}=0$.

\subsubsection{The bounds of $\vert{\langle m\rangle^{}_{\mu\tau}}\vert$}

Quite similar to the 3D profile of $\vert{\langle m\rangle^{}_{\mu\mu}}\vert$
discussed above, the upper and lower layers of
$\vert{\langle m\rangle^{}_{\mu\tau}}\vert$ are highly plain and
have no intersecting point or area, as shown in Fig.~\ref{FIG2}.
For $m^{}_1 \lesssim 0.1~{\rm eV}$,
the possibility of $\vert{\langle m\rangle^{}_{\mu\tau}}\vert=0$ has already been
excluded.

\subsection{Inverted mass ordering}

\begin{figure}[htbp]
\centering
{
\includegraphics[scale=1.34]{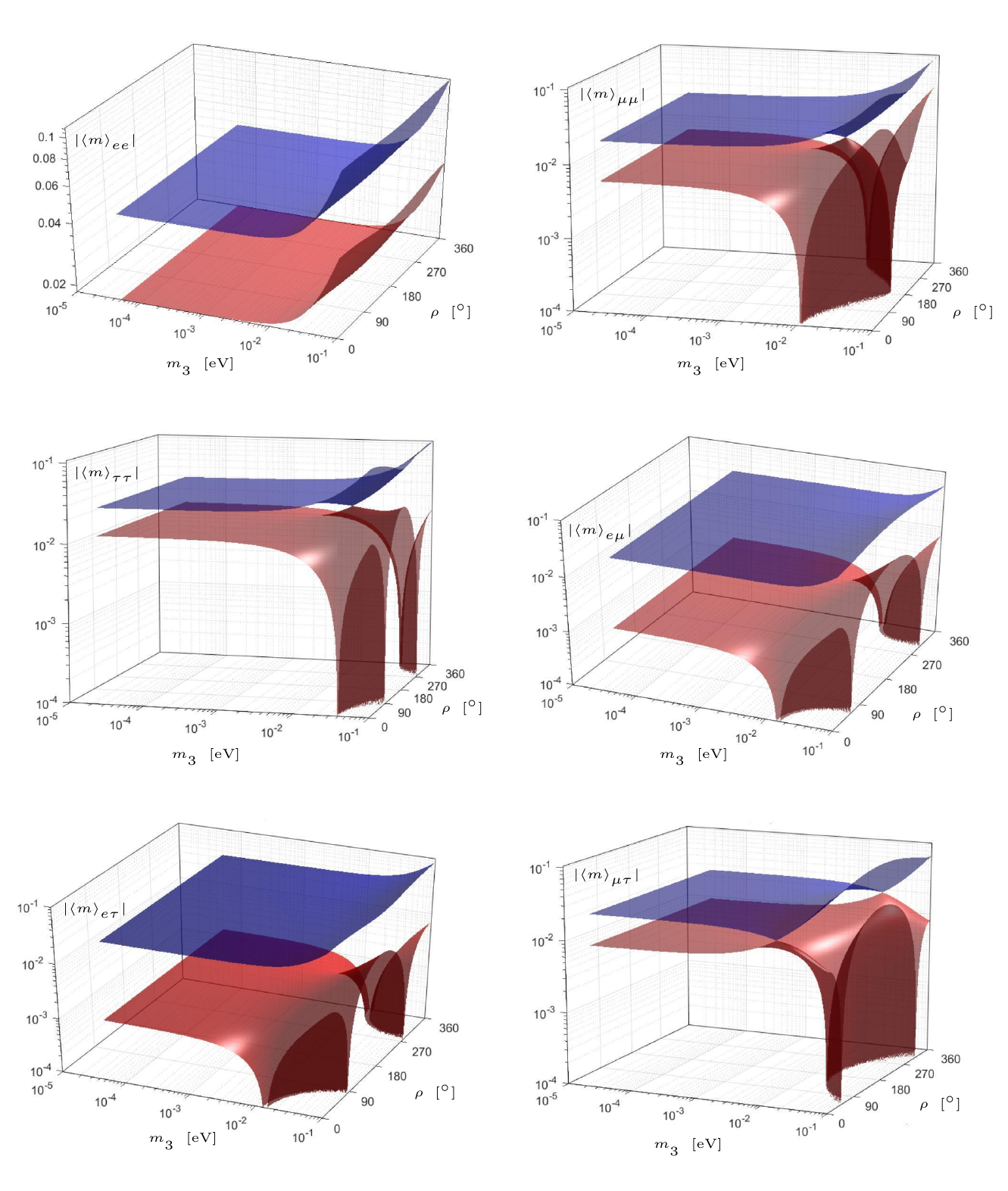}
}
\vspace{-1cm}
\caption{The 3D profile of each of the six effective Majorana neutrino masses
$|\langle m\rangle^{}_{\alpha\beta}|$ (for $\alpha, \beta = e, \mu, \tau$):
its upper (blue) and lower (red) bounds are functions of $m^{}_3$ and $\rho$
in the IMO case. Note that the two touching points between the upper and lower
layers of $|\langle m\rangle^{}_{\mu\tau}|$ are equivalent in physics, because they
correspond to $\rho = 0$ and $2\pi$.}
\label{FIG7}
\end{figure}

In the IMO case (i.e., $m^{}_3 < m^{}_1 < m^{}_2$), we have the relations
$m^{}_{1} = \sqrt{{m^2_{3}} - \Delta m^{2}_{32} - \Delta m^{2}_{21}}$
and $m^{}_{2} = \sqrt{{m^2_{3}} - \Delta m^{2}_{32}}$ with
$\Delta m^2_{21}$ and $\Delta m^2_{32}$ having been extracted from
 current neutrino oscillation data \cite{Esteban:2020cvm}.
So let us choose the smallest neutrino mass $m^{}_3$, together with the effective
Majorana phase $\rho$, to describe the profiles of $|\langle m\rangle^{}_{\alpha\beta}|$.
Since the upper and lower bounds of $|\langle m\rangle^{}_{\alpha\beta}|$
against the other Majorana phase $\sigma$ have been determined in Eqs.~(16)---(21),
here we numerically illustrate their 3D profiles in Fig.~\ref{FIG7}. More concrete
discussions are in order.

\subsubsection{The bounds of $\vert{\langle m\rangle^{}_{ee}}\vert$}

As shown in Fig.~\ref{FIG7}, the lower and upper layers of $\vert{\langle m\rangle^{}_{ee}}\vert$ are essentially flat, stable, parallel to each other
and insensitive to the changes of $\rho$;
and their magnitudes are around $0.01~{\rm eV}$ and $0.1~{\rm eV}$, respectively,
when $m^{}_3 \lesssim 0.01~{\rm eV}$ holds. These interesting features are quite
different from those salient features of $\vert{\langle m\rangle^{}_{ee}}\vert$
in the NMO case, where the parameter space of
$\vert{\langle m\rangle^{}_{ee}}\vert$ is rather tricky and even
$\vert{\langle m\rangle^{}_{ee}}\vert = 0$ is possible.

It is optimistically expected that the next-generation $0\nu 2\beta$ experiments
might be able to probe the IMO possibility of massive neutrinos if their
sensitivities to the half-lives of the decaying isotopes (such as $^{76}_{32}{\rm Ge}$,
$^{136}_{~54}{\rm Xe}$ and $^{130}_{~52}{\rm Te}$) can finally reach the level
of $10^{27}$ or even $10^{28}$ years \cite{Dolinski:2019nrj,Agostini:2018tnm}.
But a global analysis of today's neutrino oscillation data indicates that the NMO
possibility seems to be slightly favored \cite{Capozzi:2021fjo}. The next-generation
neutrino oscillation experiments, especially JUNO, DUNE and Hyper-Kamiokande
\cite{Zyla:2020zbs} will determine the neutrino mass ordering and thus greatly reduce
the uncertainties of $\vert{\langle m\rangle^{}_{\alpha\beta}}\vert$ under
discussion.

\subsubsection{The bounds of $\vert{\langle m\rangle^{}_{\mu\mu}}\vert$}

Fig.~\ref{FIG7} tells us that the profile of $\vert{\langle m\rangle^{}_{\mu\mu}}\vert$
in the IMO case is highly nontrivial as compared with that in the NMO case. Namely,
$\vert{\langle m\rangle^{}_{\mu\mu}}\vert^{}_{\rm U}$ and
$\vert{\langle m\rangle^{}_{\mu\mu}}\vert^{}_{\rm L}$ have a touching point, and
$\vert{\langle m\rangle^{}_{\mu\mu}}\vert^{}_{\rm L}$ is possible to vanish.
Taking $\vert{\langle m\rangle^{}_{\mu\mu}}\vert^{}_{\rm U} =
\vert{\langle m\rangle^{}_{\mu\mu}}\vert^{}_{\rm L} \equiv
\vert{\langle m\rangle^{}_{\mu\mu}}\vert^{}_*$, we find that the touching point
can be determined from the condition
\begin{eqnarray}
m^{}_{1} e^{\rm i\rho}\left|\left(s^{}_{12}c^{}_{23} + c^{}_{12}s^{}_{23}s^{}_{13}
e^{\rm i\delta}\right)^{2}\right| + m^{}_{3}{c^{2}_{13}}{s^{2}_{23}} = 0 \; .
\end{eqnarray}
It is then straightforward to obtain $\rho = \pi$,
\begin{eqnarray}
m^{}_3 = \frac{\displaystyle \sqrt{-\Delta m^2_{32}-\Delta m^2_{21}} \left|
\left( s^{}_{12}c^{}_{23}+c^{}_{12}s^{}_{23}s^{}_{13} e^{\rm i\delta}\right)^2
\right|}{\displaystyle \sqrt{c^4_{13} s^4_{23}- \left|\left( s^{}_{12}c^{}_{23}+c^{}_{12}s^{}_{23}s^{}_{13}
e^{\rm i\delta}\right)^2 \right|^2}} \; ,
\end{eqnarray}
and
\begin{eqnarray}
\vert{\langle m\rangle^{}_{\mu\mu}}\vert^{}_* = m^{}_2 \left|\left( c^{}_{12}c^{}_{23} - s^{}_{12}s^{}_{23}s^{}_{13} e^{\rm i\delta}\right)^2\right|
= \sqrt{m^2_{3}- \Delta m^2_{32}} \left|\left( c^{}_{12}c^{}_{23} -
s^{}_{12}s^{}_{23}s^{}_{13} e^{\rm i\delta}\right)^2\right| \; ,
\end{eqnarray}
where $\Delta m^2_{32} <0$. Numerically, we have
$\left(\rho, m^{}_{3}\right) \simeq \left(\pi, 0,0143~{\rm eV}\right)$ and
$\vert{\langle m\rangle^{}_{\mu\mu}}\vert^{}_* \simeq 0.0146 ~{\rm eV}$ for
the touching point. At this point the original Majorana phase $\xi^{}_1$ is
related to the Dirac phase $\delta$ via the relation $\xi^{}_{1} = \pi-\arg\left[\left(s^{}_{12}c^{}_{23}+c^{}_{12}s^{}_{23}s^{}_{13}
e^{\rm i\delta}\right)^{2}\right]$.

\begin{figure}
\centering
\includegraphics[scale=0.44]{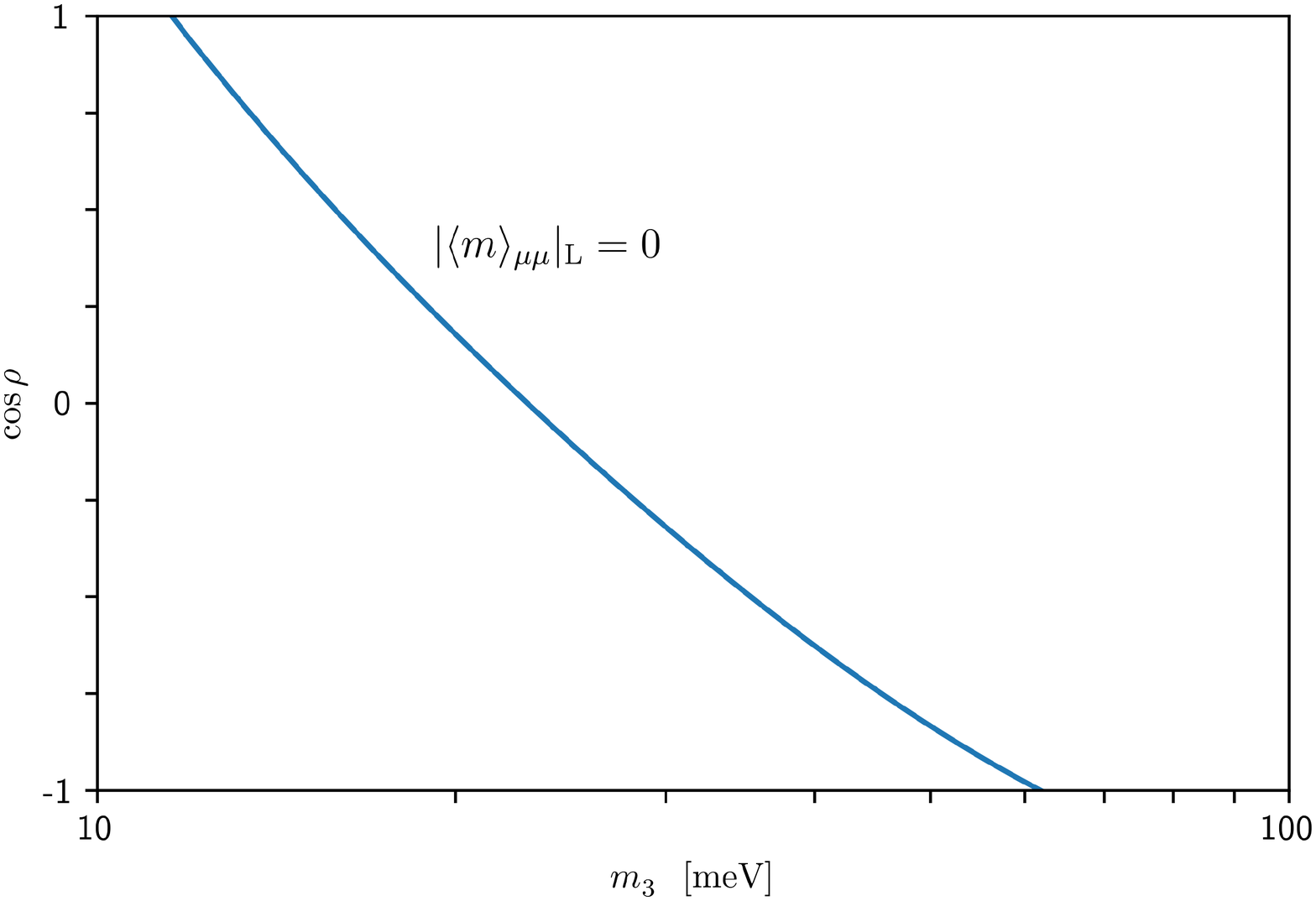}
\vspace{-0.9cm}
\caption{The correlation between $\cos \rho$ and $m^{}_{3}$ constrained by
$\vert{\langle m\rangle^{}_{\mu\mu}}\vert^{}_{\rm L}=0$ in the IMO case.}
\label{FIG8}
\end{figure}

In addition, let us consider the possibility of vanishing
$\vert{\langle m\rangle^{}_{\mu\mu}}\vert$. We find that
$\vert{\langle m\rangle^{}_{\mu\mu}}\vert^{}_{\rm L} = 0$ will be
achieved if the condition
\begin{eqnarray}
\cos \rho = \frac{\displaystyle m^2_2 \left|\left(c^{}_{12}c^{}_{23}-s^{}_{12}s^{}_{23}s^{}_{13} e^{\rm i\delta}\right)^2\right|^2 - m^2_{1}\left|\left(s^{}_{12}c^{}_{23} + c^{}_{12}s^{}_{23}s^{}_{13} e^{\rm i\delta}\right)^2\right|^2 - m^2_{3}c^4_{13}s^4_{23}}{\displaystyle 2m^{}_{1} m^{}_{3} c^2_{13}s^2_{23}\left|\left(s^{}_{12}c^{}_{23} + c^{}_{12}s^{}_{23}s^{}_{13}
e^{\rm i\delta}\right)^2\right|}
\end{eqnarray}
is satisfied. A numerical illustration of this condition is given in Fig.~\ref{FIG8},
from which one can see $0.01 ~{\rm eV} \lesssim m^{}_{3} \lesssim 0.06 ~{\rm eV}$ as
constrained by $\vert{\langle m\rangle^{}_{\mu\mu}}\vert^{}_{\rm L} = 0$.

\subsubsection{The bounds of $\vert{\langle m\rangle^{}_{\tau\tau}}\vert$}

Fig.~\ref{FIG7} shows that the 3D profile of $\vert{\langle m\rangle^{}_{\tau\tau}}\vert$
is quite similar to that of $\vert{\langle m\rangle^{}_{\mu\mu}}\vert$ in the IMO case,
implying an approximate $\mu$-$\tau$ symmetry between them.
The location of the touching point between the upper and lower layer of
$\vert{\langle m\rangle^{}_{\tau\tau}}\vert$ can be fixed by taking
$\vert{\langle m\rangle^{}_{\tau\tau}}\vert^{}_{\rm U} =
\vert{\langle m\rangle^{}_{\tau\tau}}\vert^{}_{\rm L} \equiv
\vert{\langle m\rangle^{}_{\tau\tau}}\vert^{}_*$, namely
\begin{eqnarray}
m^{}_{1}e^{\rm i\rho} \left|\left(s^{}_{12}s^{}_{23}-c^{}_{12}c^{}_{23}s^{}_{13}
e^{\rm i\delta}\right)^{2}\right| + m^{}_{3}{c^{2}_{13}}{c^{2}_{23}} = 0 \; .
\end{eqnarray}
As a result, we arrive at $\rho = \pi$,
\begin{eqnarray}
m^{}_3 = \frac{\displaystyle \sqrt{-\Delta m^2_{32}-\Delta m^2_{21}} \left|\left( s^{}_{12}s^{}_{23} - c^{}_{12}c^{}_{23}s^{}_{13} e^{\rm i\delta}\right)^2\right|}
{\displaystyle \sqrt{c^4_{13}c^4_{23}- \left|\left( s^{}_{12}s^{}_{23} - c^{}_{12}c^{}_{23}s^{}_{13} e^{\rm i\delta}\right)^2\right|^2}} \; ,
\end{eqnarray}
and
\begin{eqnarray}
\vert{\langle m\rangle^{}_{\tau\tau}}\vert^{}_* = m^{}_2\left|\left( c^{}_{12}s^{}_{23} + s^{}_{12}c^{}_{23}s^{}_{13}e^{\rm i\delta}\right)^2\right|
= \sqrt{m^2_{3}- \Delta m^2_{32}} \left|\left( c^{}_{12}s^{}_{23} +
s^{}_{12}c^{}_{23}s^{}_{13} e^{\rm i\delta}\right)^2\right| \;
\end{eqnarray}
from Eq.~(39). Numerically, $\left(\rho, m^{}_{3}\right) \simeq
\left(\pi, 0.0209 ~{\rm eV}\right)$ and $\vert{\langle m\rangle^{}_{\tau\tau}}\vert^{}_*
\simeq 0.0228 ~{\rm eV}$ for the touching point. As for the original Majorana phase
at this point,
$\xi^{}_{1} = \pi-\arg\left[\left(s^{}_{12}s^{}_{23}-c^{}_{12}c^{}_{23}s^{}_{13}
e^{\rm i\delta}\right)^{2}\right]$.

\begin{figure}
\centering
\includegraphics[scale=0.44]{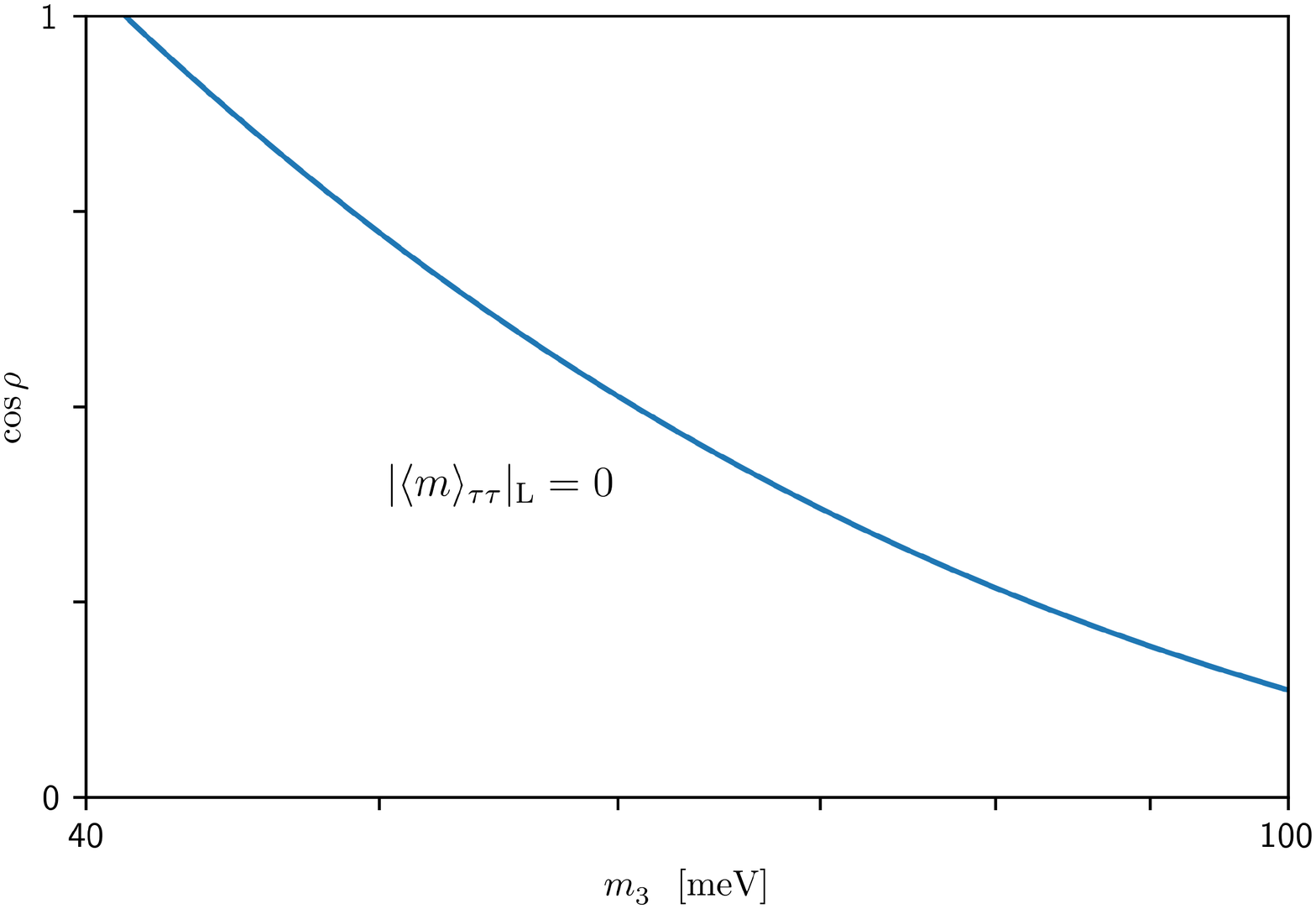}
\vspace{-0.9cm}
\caption{The correlation between $\cos \rho$ and $m^{}_{3}$ constrained by
$\vert{\langle m\rangle^{}_{\tau\tau}}\vert^{}_{\rm L}=0$ in the IMO case.}
\label{FIG9}
\end{figure}

Now that Eqs.~(39)---(41) can be directly achieved from Eqs.~(35)---(37) by making
the replacements $c^{}_{23} \to s^{}_{23}$ and $s^{}_{23} \to -c^{}_{23}$, we
may simply write out the condition for
$\vert{\langle m\rangle^{}_{\tau\tau}}\vert^{}_{\rm L} = 0$ with the help of
Eq.~(38) by making the same replacements. That is,
$\vert{\langle m\rangle^{}_{\tau\tau}}\vert^{}_{\rm L} = 0$ will hold if
\begin{eqnarray}
\cos \rho = \frac{\displaystyle m^2_2 \left|\left(c^{}_{12}s^{}_{23}+s^{}_{12}c^{}_{23}
s^{}_{13} e^{\rm i\delta}\right)^2\right|^2 - m^2_{1}\left|\left(s^{}_{12}s^{}_{23} - c^{}_{12}c^{}_{23}s^{}_{13} e^{\rm i\delta}\right)^2\right|^2 - m^2_{3}c^4_{13}c^4_{23}}{\displaystyle 2m^{}_{1} m^{}_{3} c^2_{13}c^2_{23}\left|\left(s^{}_{12}s^{}_{23} - c^{}_{12}c^{}_{23}s^{}_{13}
e^{\rm i\delta}\right)^2\right|} \;
\end{eqnarray}
is satisfied. This condition is formally the same as that obtained in Eq.~(26)
for the NMO case. A numerical illustration of Eq.~(42) is presented in
Fig.~\ref{FIG9}. We immediately see that $m^{}_{3} \gtrsim 0.04 ~{\rm eV}$ is
required for $\vert{\langle m\rangle^{}_{\tau\tau}}\vert^{}_{\rm L}$ to vanish in
the IMO case.

\subsubsection{The bounds of $\vert{\langle m\rangle^{}_{e\mu}}\vert$}

Comparing between the 3D profile of $\vert{\langle m\rangle^{}_{e\mu}}\vert$ in the
NMO case shown in Fig.~\ref{FIG2} and that in the IMO case shown in Fig.~\ref{FIG7},
we find that the latter is somewhat less structured (e.g., its upper and lower layers
have no touching or intersecting point in the given parameter space). But in both
cases $\vert{\langle m\rangle^{}_{e\mu}}\vert^{}_{\rm L}$ is possible to vanish.

\begin{figure}
\centering
\includegraphics[scale=0.44]{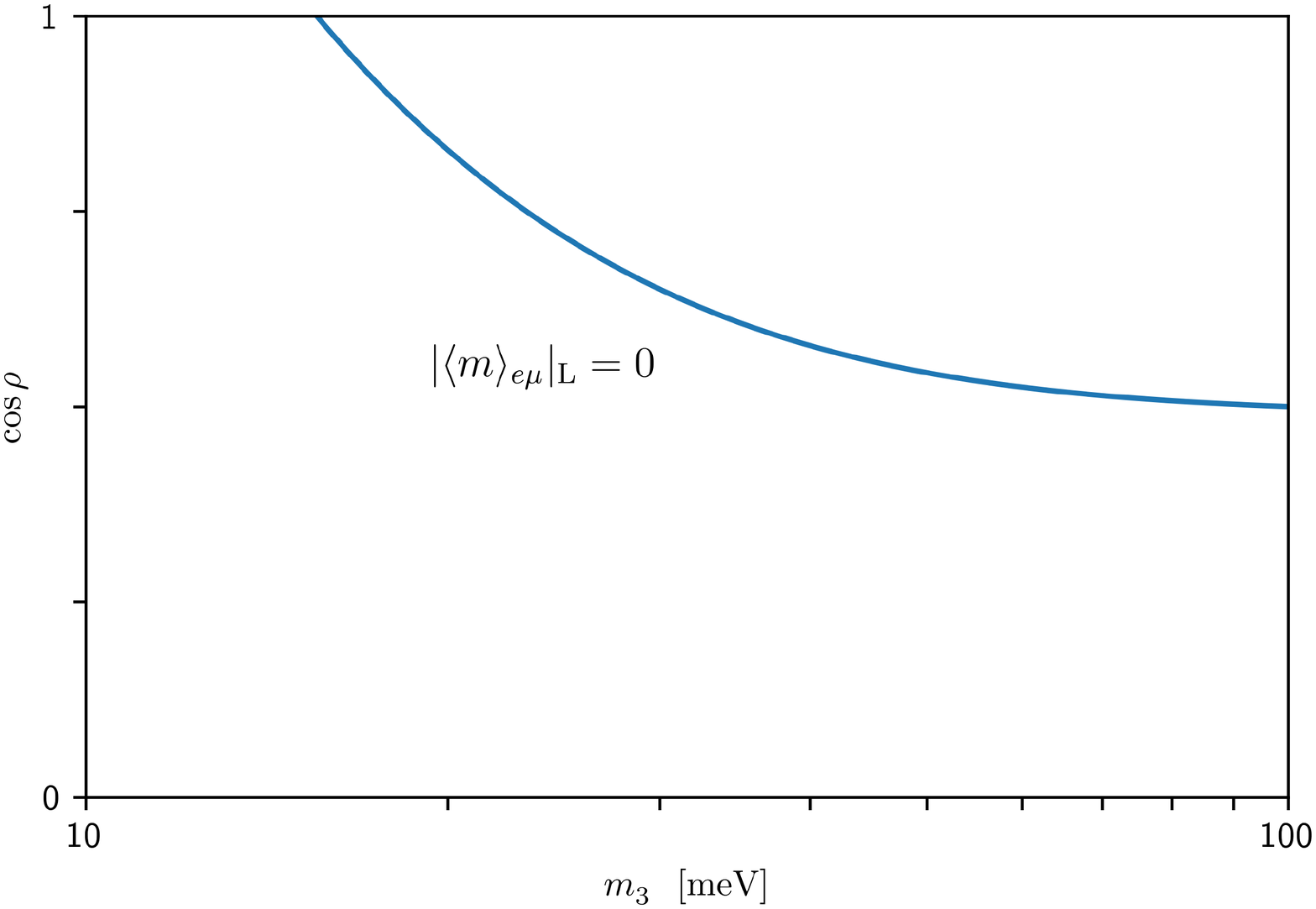}
\vspace{-0.9cm}
\caption{The correlation between $\cos \rho$ and $m^{}_{3}$ constrained by
$\vert{\langle m\rangle^{}_{e\mu}}\vert^{}_{\rm L}=0$ in the IMO case.}
\label{FIG10}
\end{figure}

The specific condition for $\vert{\langle m\rangle_{e\mu}}\vert^{}_{\rm L}=0$
has been given in Eq.~(30), but now it is subject to the IMO case and thus
yields a different curve as illustrated by Fig.~\ref{FIG10}. One can see
that $\vert{\langle m\rangle^{}_{e\mu}}\vert^{}_{\rm L}=0$ requires the smallest
neutrino mass $m^{}_3$ to lie in the range of $m^{}_3 \gtrsim 0.01 ~{\rm eV}$.

\subsubsection{The bounds of $\vert{\langle m\rangle^{}_{e\tau}}\vert$}

As shown in Fig.~\ref{FIG7}, the 3D profile of $\vert{\langle m\rangle^{}_{e\tau}}\vert$
with its upper and lower layers exhibits a striking similarity to that of
$\vert{\langle m\rangle^{}_{e\mu}}\vert$ in the IMO case. This interesting feature
demonstrates the existence of an approximate $\mu$-$\tau$ symmetry between these
two effective Majorana neutrino masses as supported by current neutrino oscillation
data \cite{Xing:2015fdg}.

\begin{figure}
\centering
\includegraphics[scale=0.44]{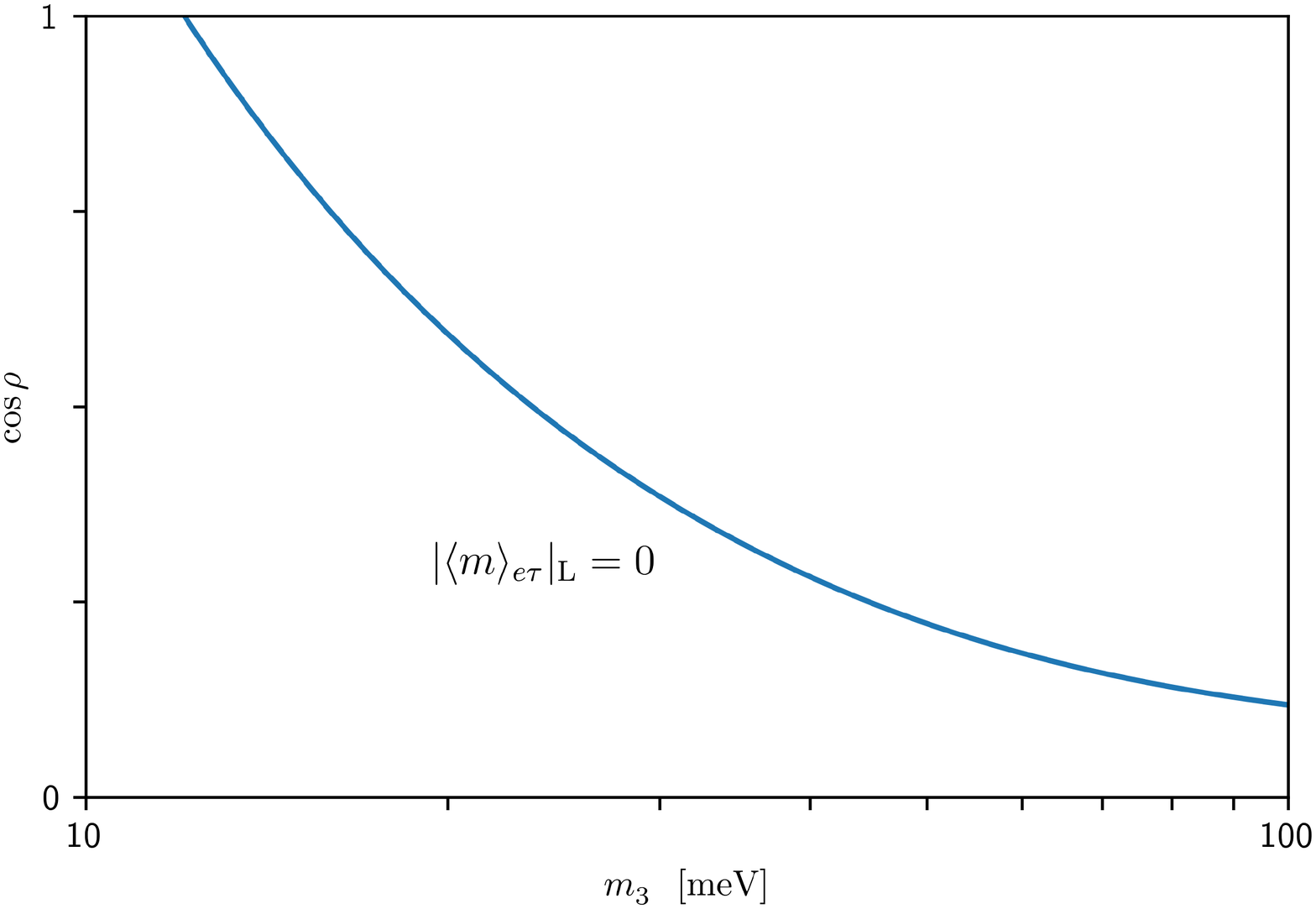}
\vspace{-0.9cm}
\caption{The correlation between $\cos \rho$ and $m^{}_{3}$ constrained by
$\vert{\langle m\rangle^{}_{e\tau}}\vert^{}_{\rm L}=0$ in the IMO case.}
\label{FIG11}
\end{figure}

Similarly, $\vert{\langle m\rangle^{}_{e\tau}}\vert^{}_{\rm L} = 0$ will hold
if the condition presented in Eq.~(34) is satisfied in the IMO case. The corresponding
numerical illustration of this condition is shown in Fig.~\ref{FIG11}, which
is quite similar to Fig.~\ref{FIG10} as a direct consequence of the approximate $\mu$-$\tau$
symmetry mentioned above. It is obvious that $m^{}_3 \gtrsim 0.01~{\rm eV}$ is
required to obtain $\vert{\langle m\rangle^{}_{e\tau}}\vert^{}_{\rm L}=0$.

\subsubsection{The bounds of $\vert{\langle m\rangle^{}_{\mu\tau}}\vert$}

As shown in Fig.~\ref{FIG7}, the upper and lower layers of
$\vert{\langle m\rangle^{}_{\mu\tau}}\vert$ have a touching point determined
by the condition $\vert{\langle m\rangle^{}_{\mu\tau}}\vert^{}_{\rm U} =
\vert{\langle m\rangle^{}_{\mu\tau}}\vert^{}_{\rm L} \equiv
\vert{\langle m\rangle^{}_{\mu\tau}}\vert^{}_*$. The latter means
\begin{eqnarray}
-m^{}_{1}e^{\rm i\rho}\left|\left(s^{}_{12}c^{}_{23}+c^{}_{12}s^{}_{23}s^{}_{13}
e^{\rm i\delta}\right)\left(s^{}_{12}s^{}_{23}-c^{}_{12}c^{}_{23}s^{}_{13}
e^{\rm i\delta}\right) \right| + m^{}_{3}c^{2}_{13}c^{}_{23}s^{}_{23} = 0 \; ,
\end{eqnarray}
from which we arrive at $\rho = 0$ (or $2\pi$),
\begin{eqnarray}
m^{}_3 = \frac{\displaystyle \sqrt{-\Delta m^2_{32} - \Delta m^2_{21}}
\left|\left(s^{}_{12}c^{}_{23}+c^{}_{12}s^{}_{23}s^{}_{13}
e^{\rm i\delta}\right)\left(s^{}_{12}s^{}_{23} - c^{}_{12}c^{}_{23}s^{}_{13} e^{\rm i\delta}\right)\right|}
{\displaystyle \sqrt{c^4_{13}c^2_{23}s^2_{23} - \left|\left(s^{}_{12}c^{}_{23}+c^{}_{12}s^{}_{23}s^{}_{13}
e^{\rm i\delta}\right)\left(s^{}_{12}s^{}_{23}-c^{}_{12}c^{}_{23}s^{}_{13} e^{\rm i\delta}\right)\right|^2}} \; ,
\end{eqnarray}
and thus
\begin{eqnarray}
\vert{\langle m\rangle^{}_{\mu\tau}}\vert^{}_* \hspace{-0.2cm} & = & \hspace{-0.2cm}
m^{}_{2}\left|\left(c^{}_{12} c^{}_{23} - s^{}_{12} s^{}_{23} s^{}_{13}
e^{{\rm i}\delta}\right)\left(c^{}_{12} s^{}_{23} + s^{}_{12} c^{}_{23} s^{}_{13}
e^{{\rm i}\delta}\right)\right|
\nonumber \\
\hspace{-0.2cm} & = & \hspace{-0.2cm}
\sqrt{m^2_3-\Delta m^2_{32}} \left|\left(c^{}_{12} c^{}_{23}
- s^{}_{12} s^{}_{23} s^{}_{13} e^{{\rm i}\delta}\right)\left(c^{}_{12} s^{}_{23} +
s^{}_{12} c^{}_{23} s^{}_{13} e^{{\rm i}\delta}\right) \right| \; . \hspace{0.5cm}
\end{eqnarray}
Numerically, we obtain $(\rho, m^{}_{3}) \simeq (0, 0.0172 ~{\rm eV})$ and
$\vert{\langle m\rangle^{}_{\mu\tau}}\vert^{}_* \simeq 0.0182 ~{\rm eV}$ for
the touching point. At this point we are left with
$\xi^{}_{1} = -\arg\left[\left(s^{}_{12}c^{}_{23}+c^{}_{12}s^{}_{23}s^{}_{13}
e^{\rm i\delta}\right)\left(s^{}_{12}s^{}_{23}-c^{}_{12}c^{}_{23}s^{}_{13}
e^{\rm i\delta}\right)\right]$ which links the original Majorana phase $\xi^{}_1$
to the Dirac phase $\delta$.

\begin{figure}
\centering
\includegraphics[scale=0.5]{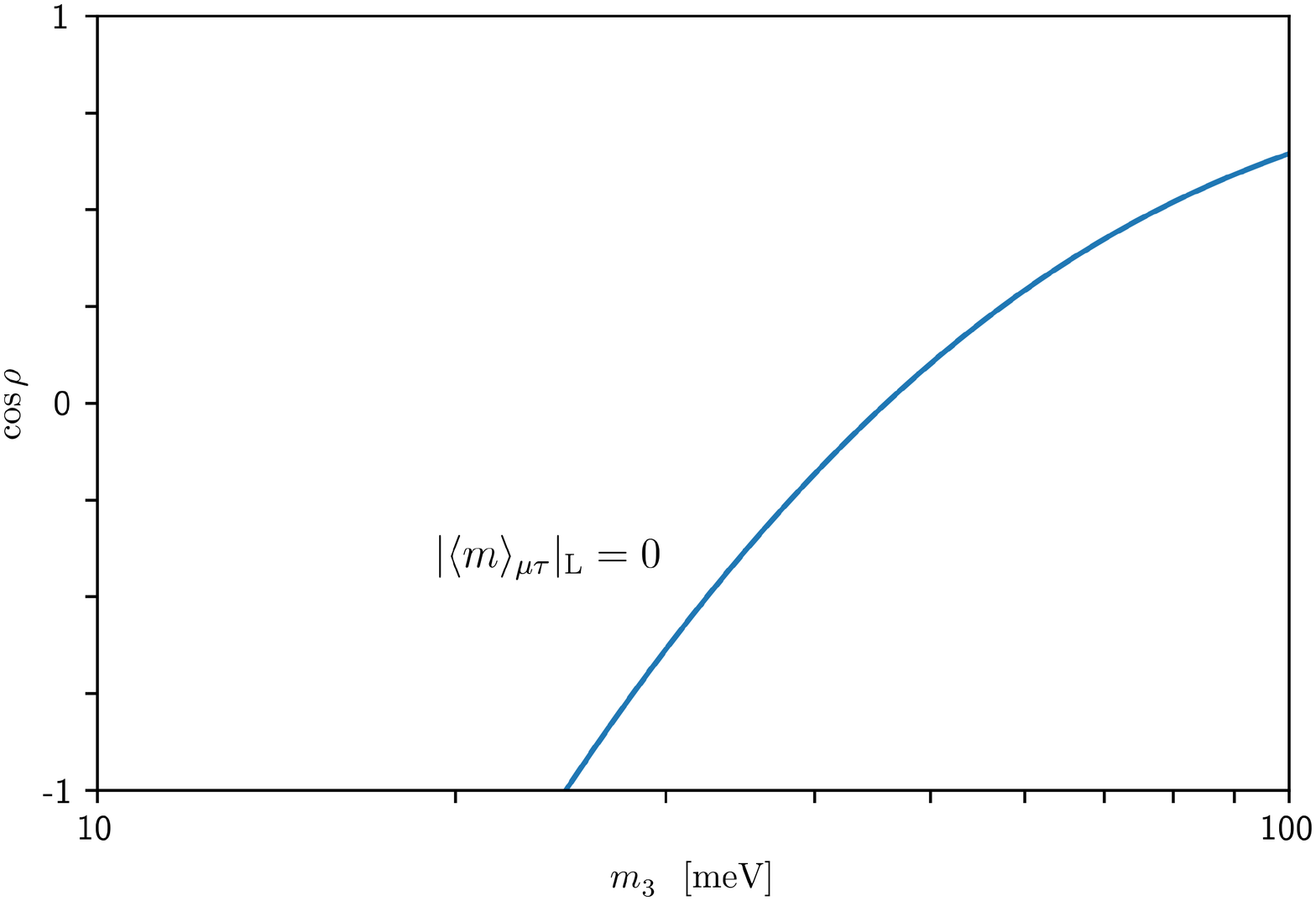}
\vspace{-0.9cm}
\caption{The correlation between $\cos \rho$ and $m^{}_{3}$ constrained by
$\vert{\langle m\rangle^{}_{e\tau}}\vert^{}_{\rm L}=0$ in the IMO case.}
\label{FIG12}
\end{figure}

On the other hand, $\vert{\langle m\rangle^{}_{\mu\tau}}\vert^{}_{\rm L} = 0$ will
hold if the condition
\begin{eqnarray}
\cos \rho \hspace{-0.2cm} & = & \hspace{-0.2cm}
\frac{\displaystyle m^2_{3}c^4_{13}c^2_{23}s^2_{23} + m^2_{1}\left|\left(s^{}_{12}c^{}_{23}+c^{}_{12}s^{}_{23}s^{}_{13}
e^{\rm i\delta}\right) \left(s^{}_{12}s^{}_{23}-c^{}_{12}c^{}_{23}s^{}_{13}
e^{\rm i\delta}\right)\right|^2}
{\displaystyle 2m^{}_1 m^{}_3 c^2_{13}c^{}_{23}s^{}_{23} \left|\left(s^{}_{12}c^{}_{23}+c^{}_{12}s^{}_{23}s^{}_{13} e^{\rm i\delta}
\right) \left(s^{}_{12}s^{}_{23}-c^{}_{12}c^{}_{23}s^{}_{13} e^{\rm i\delta}
\right)\right|}
\nonumber \\
\hspace{-0.2cm} & & \hspace{-0.2cm}
- \frac{\displaystyle m^2_{2}\left|\left(c^{}_{12} c^{}_{23} -
s^{}_{12} s^{}_{23} s^{}_{13} e^{{\rm i}\delta}\right)\left(c^{}_{12} s^{}_{23} + s^{}_{12}
c^{}_{23} s^{}_{13} e^{{\rm i}\delta}\right)\right|^2}
{\displaystyle 2m^{}_1 m^{}_3 c^2_{13}c^{}_{23}s^{}_{23} \left|\left(s^{}_{12}c^{}_{23}+c^{}_{12}s^{}_{23}s^{}_{13} e^{\rm i\delta}
\right)\left(s^{}_{12}s^{}_{23}-c^{}_{12}c^{}_{23}s^{}_{13} e^{\rm i\delta}
\right)\right|} \;  \hspace{0.5cm}
\end{eqnarray}
is satisfied. This condition is numerically illustrated in Fig~\ref{FIG12},
where one can see $m^{}_3 \gtrsim 0.02~{\rm eV}$ should hold as required by
$\vert{\langle m\rangle^{}_{\mu\tau}}\vert^{}_{\rm L}=0$.

\section{Some further discussions}

So far we have discussed the 3D profiles of all the six independent effective Majorana
neutrino masses $|\langle m\rangle^{}_{\alpha\beta}|$ with respect to the effective
Majorana phase $\rho$ and the smallest neutrino mass $m^{}_1$ (or $m^{}_3$) in the
NMO (or IMO) case, where the other effective Majorana phase $\sigma$ has been eliminated
by determining the upper and lowers bounds of each $|\langle m\rangle^{}_{\alpha\beta}|$
against its corresponding $\sigma$. Such a phenomenological
strategy is of course imperfect, but it can help a lot in simplifying our analytical
and numerical descriptions of $|\langle m\rangle^{}_{\alpha\beta}|$.
Compared with the conventional 2D plots of $|\langle m\rangle^{}_{\alpha\beta}|$
against $m^{}_1$ or $m^{}_3$, where all the CP-violating phases of the PMNS matrix
$U$ are allowed to vary in their respective ranges, this 3D mapping of
$|\langle m\rangle^{}_{\alpha\beta}|$ is at least advantageous in the aspects of
unraveling the dependence of each effective Majorana mass on its phase parameter
$\rho$ and giving us a ball-park feeling of the bulk of its 3D parameter space.

In this respect we are going to further discuss the following three issues:
(1) the possibility of a four-dimensional (4D) description of
$|\langle m\rangle^{}_{\alpha\beta}|$ with
respect to $m^{}_1$ (or $m^{}_3$), $\rho$ and $\sigma$;
(2) the arbitrariness in defining the two effective Majorana phases of
$|\langle m\rangle^{}_{\alpha\beta}|$; and (3) possible texture zeros and (or) an
approximate $\mu$-$\tau$ symmetry hidden in the effective Majorana neutrino mass matrix
$M^{}_\nu$ as revealed in our analyses made above.

\subsection{The 4D plots of $|\langle m\rangle^{}_{\alpha\beta}|$}

\begin{figure}[htbp]
\centering
{
\includegraphics[width=17.5cm]{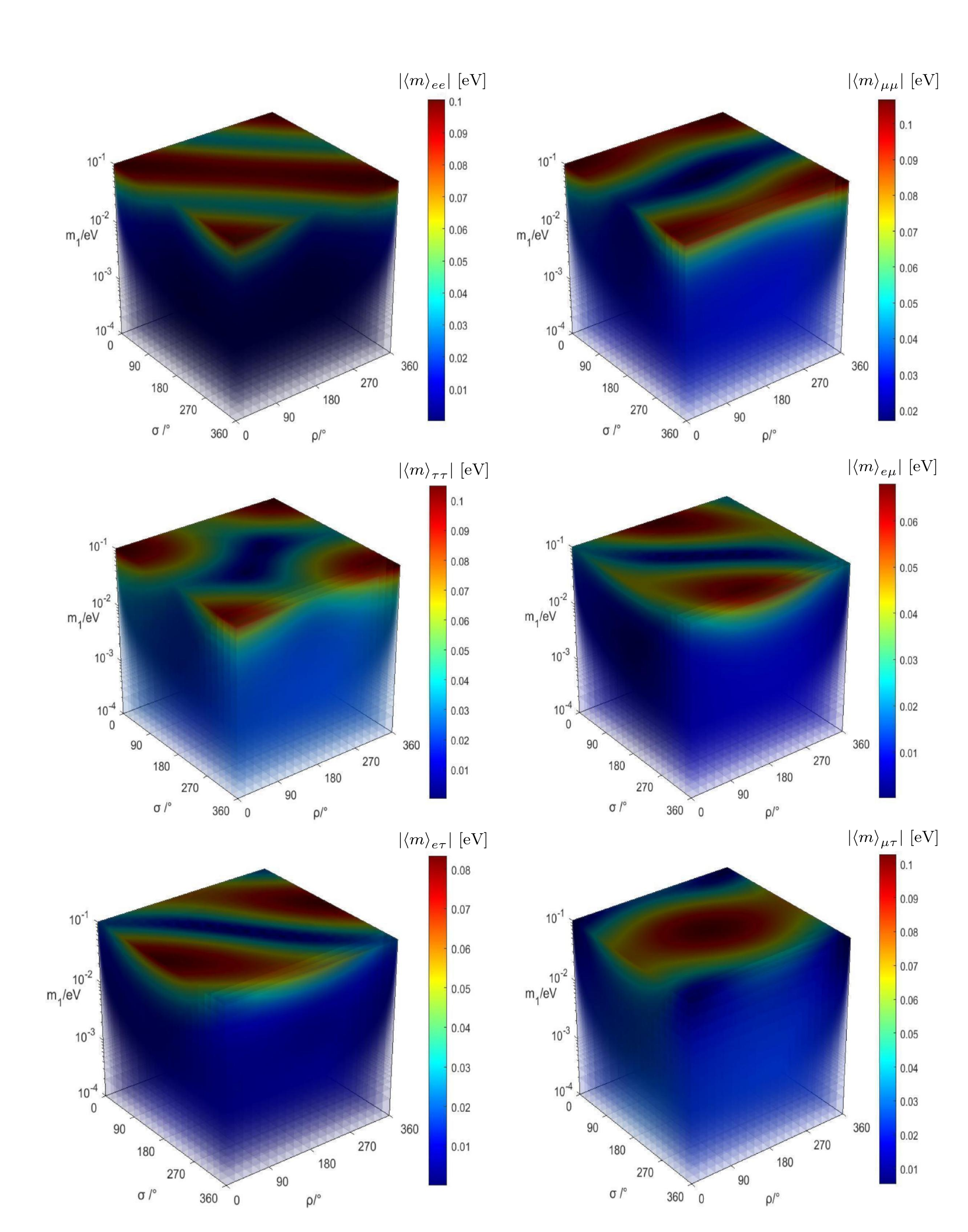}
}
\vspace{-0.8cm}
\caption{The 4D plot of each of the six effective Majorana neutrino masses
$|\langle m\rangle^{}_{\alpha\beta}|$ (for $\alpha, \beta = e, \mu, \tau$)
with respect to $m^{}_1$, $\rho$ and $\sigma$ in the NMO case.}
\label{FIG13}
\end{figure}

\begin{figure}[htbp]
\centering
{
\includegraphics[width=18cm]{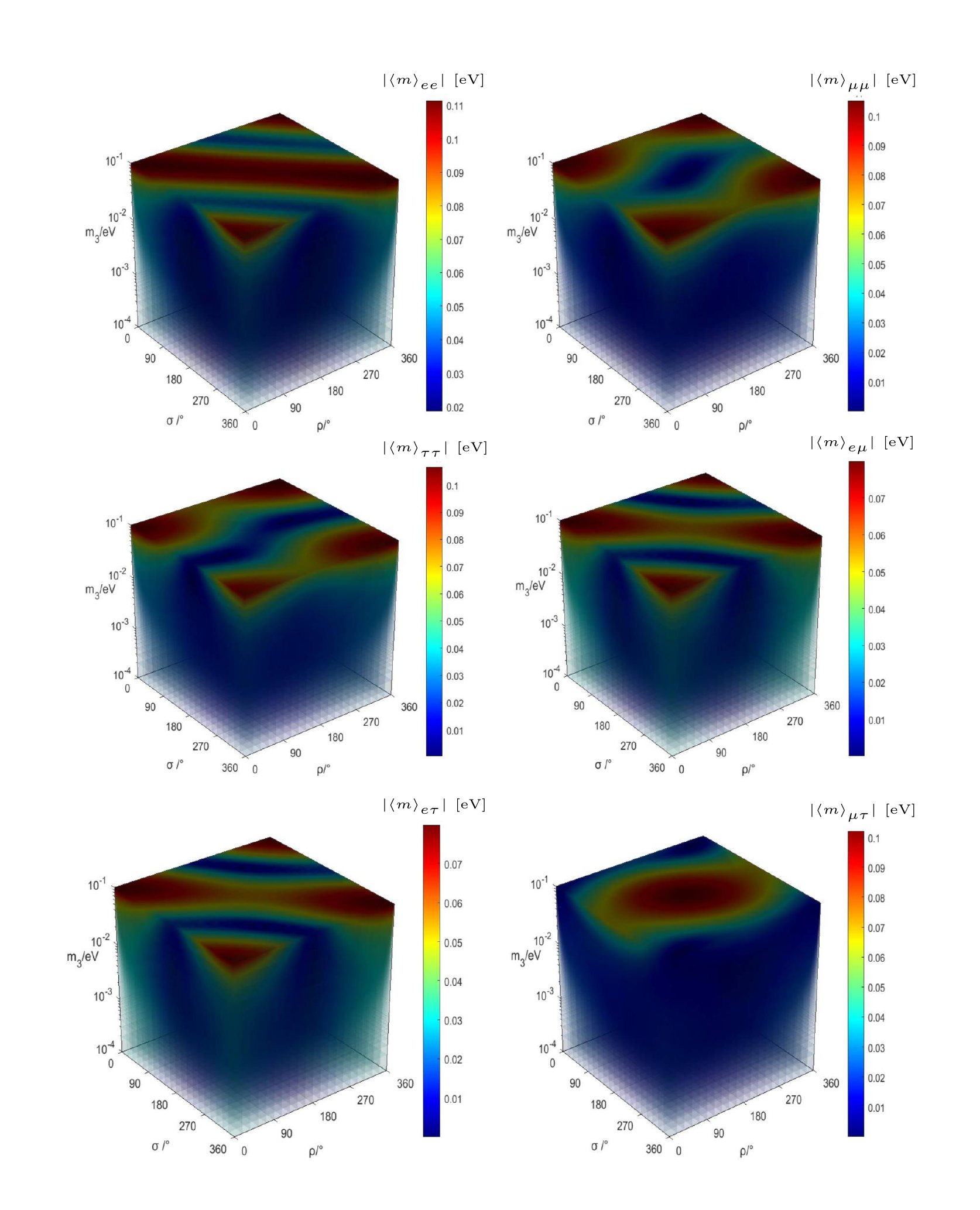}
}
\vspace{-1.2cm}
\caption{The 4D plot of each of the six effective Majorana neutrino masses
$|\langle m\rangle^{}_{\alpha\beta}|$ (for $\alpha, \beta = e, \mu, \tau$)
with respect to $m^{}_3$, $\rho$ and $\sigma$ in the IMO case.}
\label{FIG14}
\end{figure}

Fig.~\ref{FIG13} shows the 4D plots of $|\langle m\rangle^{}_{\alpha\beta}|$
with respect to the three free parameters $m^{}_1$, $\rho$ and $\sigma$ in the NMO case,
and Fig.~\ref{FIG14} illustrates the similar plots as functions of $m^{}_3$, $\rho$ and
$\sigma$ in the IMO case. One can clearly see that the bigger $m^{}_1$ (or $m^{}_3$) goes,
the bigger $|\langle m\rangle^{}_{\alpha\beta}|$ will be.
Judging from the color bar code attached to every plot in Figs.~\ref{FIG13} and
\ref{FIG14}, one may roughly figure out the allowed range of
$|\langle m\rangle^{}_{\alpha\beta}|$. In the NMO case, for instance, it is the matrix
element $\vert{\langle m\rangle^{}_{\mu\mu}}\vert$ that has a lower bound of
${\cal O} (0.01)$ eV no matter how small $m^{}_{1}$ is, while all the other five
matrix elements are possible to vanish. As for the IMO case, one may similarly
observe that the matrix element $\vert{\langle m\rangle^{}_{ee}}\vert$ has no
chance to vanish, but the other five matrix elements are possible to vanish.

We have to admit that the 3D mapping of $\vert{\langle m\rangle^{}_{\alpha\beta}}\vert$
presented in Fig.~\ref{FIG2} or Fig.~\ref{FIG7}
is much simpler and more transparent than its 4D analogue shown in Fig.~\ref{FIG13} or
Fig.~\ref{FIG14}. That is why we have focused on the 3D profiles of
$\vert{\langle m\rangle^{}_{\alpha\beta}}\vert$ and their salient features in this work.

\subsection{On the effective Majorana phases of $|\langle m\rangle^{}_{\alpha\beta}|$}

\begin{figure}[htbp]
\centering
{
\includegraphics[scale=1.25]{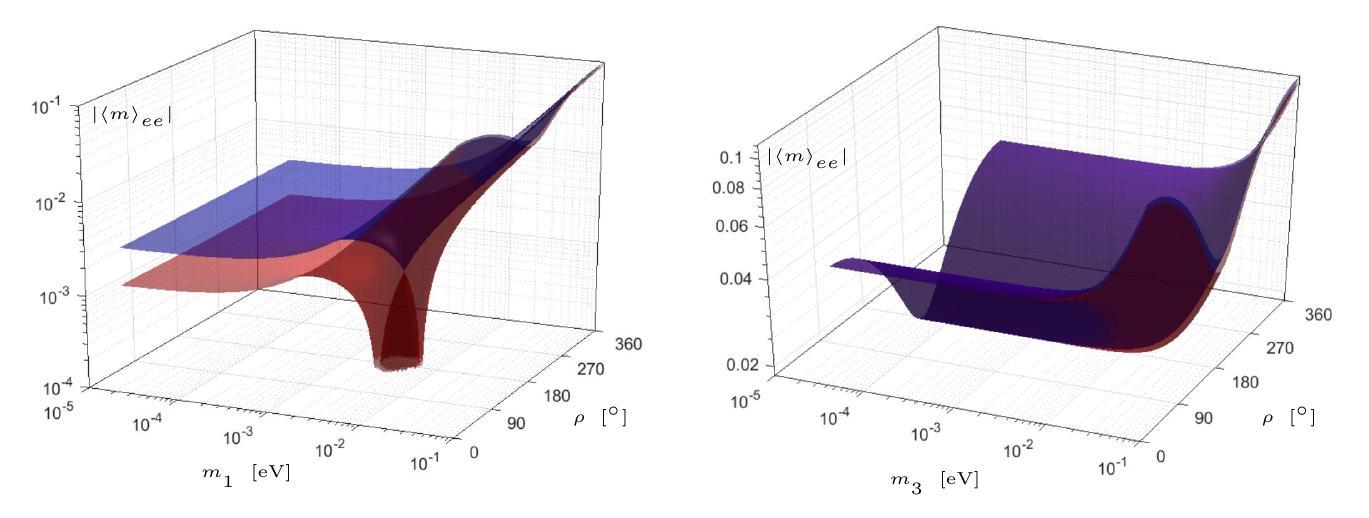}
}
\vspace{-0.9cm}
\caption{The 3D profile of $\vert{\langle m\rangle^{}_{ee}}\vert$ in either the NMO
case (left panel) or the IMO case (right panel) based on the phase assignment made in
Eq.~(47): its upper (blue) and lower (red) bounds are functions of $m^{}_1$ (or $m^{}_3$)
and $\rho$.}
\label{FIG15}
\end{figure}

In section 2.1 we have assigned the effective Majorana phases $\rho$ and $\sigma$
to the terms of $\vert{\langle m\rangle^{}_{\alpha\beta}}\vert$ that are associated
respectively with $m^{}_1$ and $m^{}_2$. The upper and lower bounds of
$\vert{\langle m\rangle^{}_{\alpha\beta}}\vert$ with respect to $\sigma$ are accordingly
determined in section 2.2. Of course, there is always a degree of
arbitrariness about such a phase assignment. In Refs.~\cite{Xing:2015zha,Xing:2016ymd,Cao:2019hli}, for example,
$\rho$ and $\sigma$ were assigned to the components of $\vert{\langle m\rangle^{}_{ee}}\vert$
that are proportional respectively to $m^{}_1$ and $m^{}_3$. This different phase
assignment leads us to $\rho=\xi^{}_{1}-\xi^{}_{2}$ and $\sigma= -\xi^{}_{2}- 2\delta$
for $\vert{\langle m\rangle^{}_{ee}}\vert$,
and the upper and lower limits of $\vert{\langle m\rangle^{}_{ee}}\vert$ with respect
to $\sigma$ turn out to be \cite{Xing:2016ymd}
\begin{eqnarray}
\vert{\langle m\rangle^{}_{ee}}\vert^{}_{\rm U,L} = \Biggl\lvert
\left| m^{ee}_{1} e^{{\rm i}\rho} + m^{ee}_{2}\right| \pm m^{ee}_{3}\Biggr\rvert \; ,
\end{eqnarray}
where $m^{ee}_1 = m^{}_1 c^2_{12} c^2_{13}$, $m^{ee}_2 = m^{}_2 s^2_{12} c^2_{13}$
and $m^{ee}_3 = m^{}_3 s^2_{13}$ have been defined just above Eq.~(16). In this case
the 3D profile of $\vert{\langle m\rangle^{}_{ee}}\vert$ for either the NMO or
the IMO is shown in Fig.~\ref{FIG15}, a result consistent with the one
obtained previously in Refs.~\cite{Xing:2016ymd,Cao:2019hli}.

Comparing the left panel of Fig.~\ref{FIG15} with the 3D profile of
$\vert{\langle m\rangle^{}_{ee}}\vert$ shown in Fig.~\ref{FIG2} in the MNO case,
one can see that the former possesses a bullet-like structure but the latter has
a crack-like structure. The reason for this remarkable difference is simply due to
the difference between the phase assignment made in Eq.~(47) and that chosen in
Eq.~(16). Similarly, the 3D profile of $\vert{\langle m\rangle^{}_{ee}}\vert$
in the IMO case is also sensitive to how to define the two effective Majorana
phases $\rho$ and $\sigma$.

Note that we have adopted the standard parametrization of the PMNS matrix $U$ in
this work, and such a choice is most favored for the study of
$\vert{\langle m\rangle^{}_{ee}}\vert$ because it makes the analytical expression
of $\vert{\langle m\rangle^{}_{ee}}\vert$ sufficiently simplified. For the study of
a given $\vert{\langle m\rangle^{}_{\alpha\beta}}\vert$, one may in principle
choose the most appropriate Euler-like parametrization of $U$ to maximally
simplify the analytical result of $\vert{\langle m\rangle^{}_{\alpha\beta}}\vert$.
But when all the six matrix elements of $M^{}_\nu$ are under discussion, just as
what we are doing, it is better to fix the parametrization of $U$ and the
phase assignment of $\vert{\langle m\rangle^{}_{\alpha\beta}}\vert$ so as to make
a comparison between any two matrix elements possible and meaningful. Our present
work represents the first attempt of this kind.

\subsection{On the $\mu$-$\tau$ symmetry and texture zeros}

We have pointed out that the 3D profiles of $\vert{\langle m\rangle^{}_{\mu\mu}}\vert$
and $\vert{\langle m\rangle^{}_{\tau\tau}}\vert$ in Fig.~\ref{FIG7} are quite similar
to each other, so are the profiles of
$\vert{\langle m\rangle^{}_{e\mu}}\vert$ and $\vert{\langle m\rangle^{}_{e\tau}}\vert$
in the IMO case. In Fig.~\ref{FIG2}, there is also a similarity between the profiles
of $\vert{\langle m\rangle^{}_{e\mu}}\vert$ and $\vert{\langle m\rangle^{}_{e\tau}}\vert$
in the NMO case, although it is not as impressive as in the IMO case. 
The straightforward reason for such similarities is because of
the existence of an approximate $\mu$-$\tau$ reflection symmetry for the matrix
elements of $M^{}_\nu$, which would become exact if $\theta^{}_{23} = \pi/4$
and $\delta = 3\pi/2$ exactly held \cite{Xing:2015fdg}.

One may wonder to what extent the 3D profiles of
$\vert{\langle m\rangle^{}_{\alpha\beta}}\vert$ shown in Fig.~\ref{FIG2} (NMO) and
Fig.~\ref{FIG7} (IMO) will change if the exact $\mu$-$\tau$ reflection symmetry is imposed
on $M^{}_\nu$. In this special case $\langle m\rangle^{}_{ee} = \langle m\rangle^{*}_{ee}$,
$\langle m\rangle^{}_{e\mu} = \langle m\rangle^{*}_{e\tau}$,
$\langle m\rangle^{}_{\mu\mu} = \langle m\rangle^{*}_{\tau\tau}$ and
$\langle m\rangle^{}_{\mu\tau} = \langle m\rangle^{*}_{\mu\tau}$ hold
\cite{Xing:2015fdg}, and thus the 3D profiles of
$\vert{\langle m\rangle^{}_{e\mu}}\vert$ and $\vert{\langle m\rangle^{}_{e\tau}}\vert$
(or those of $\vert{\langle m\rangle^{}_{\mu\mu}}\vert$ and
$\vert{\langle m\rangle^{}_{\tau\tau}}\vert$) should be exactly equal.
Fig.~\ref{FIG7} tells us that such equalities can approximately be seen, simply because
in this IMO case the input values $\theta^{}_{23} = 49.3^\circ$ and $\delta = 286^\circ$
deviate only slightly from their corresponding values in the $\mu$-$\tau$
reflection symmetry limit. In contrast, such equalities do not so
clearly show up in Fig.~\ref{FIG2} in the NMO case, where the input value
$\theta^{}_{23} = 49.0^\circ$ is quite close to $\pi/4$ but the input value
$\delta = 195^\circ$ deviates significantly from $3\pi/2$. In this connection a 
preliminary illustration of the $\mu$-$\tau$ reflection symmetry breaking effects 
have been given in Ref.~\cite{Xing:2017cwb} with the help of the 2D profiles of
$\vert{\langle m\rangle^{}_{\alpha\beta}}\vert$. Here we do not repeat a 
similar illustration of this kind in the 3D mapping scheme, since it is not
justified that fixing $\theta^{}_{23} = \pi/4$ and $\delta = 3\pi/2$ 
can still be consistent with inputting the best-fit values or $3\sigma$ intervals of 
the other neutrino oscillation parameters in such a numerical exercise.  

Different from the popular discrete flavor symmetry approach in exploring or
understanding the flavor structure of massive
neutrinos~\cite{Xing:2019vks,Feruglio:2021sir}, which can help
establish some simple equalities or linear correlations between the matrix elements
of $M^{}_\nu$, the existence of texture zeros in $M^{}_\nu$ is another phenomenological
way to reduce its free parameters and thus enhance its predictability. In this
connection the two-zero textures of $M^{}_\nu$ are most interesting and have attracted a
lot of attention (see, e.g., Refs.~\cite{Frampton:2002yf,Xing:2002ta,Xing:2002ap,Fritzsch:2011qv,Zhou:2015qua}).
As one can see from Fig.~\ref{FIG2} in the NMO case or Fig.~\ref{FIG7} in the IMO case,
$\vert{\langle m\rangle^{}_{\alpha\beta}}\vert^{}_{\rm L} = 0$ just means that a
texture zero is present.

Since our numerical analysis has been done by just inputting the best-fit values of
relevant neutrino oscillation parameters, the obtained 3D profiles of $\vert{\langle m\rangle^{}_{\alpha\beta}}\vert$ can exhibit their salient features but cannot
cover the whole parameter space. So we do not claim the coexistence of any two
texture zeros of $M^{}_\nu$ in this work, in order to avoid a possible misinterpretation
of the numerical results. A further study along this line of thought will be
done elsewhere. 

Finally, it is worth mentioning that some systematic studies of the entries
of $M^{}_\nu$ have been done
analytically, semi-analytically or numerically in the literature (see, e.g.,
Refs.~\cite{Frigerio:2002rd,Merle:2006du,Grimus:2012ii}). Here our 3D mapping
of $\vert{\langle m\rangle^{}_{\alpha\beta}}\vert$ provides a complementary
approach for exploring the parameter space of each entry of $M^{}_\nu$, or
equivalently the flavor structure of massive Majorana neutrinos, at low energies.
Of course, a similar 3D profiles of $\vert{\langle m\rangle^{}_{\alpha\beta}}\vert$
can be achieved at a superhigh energy scale after taking into account the 
renormalization-group running effects \cite{Ohlsson:2013xva}.

\section{Concluding remarks}

Everyone knows that a convincing quantitative model of neutrino masses has been
lacking, although many interesting attempts have been made in the past decades
\cite{Witten:2000dt}. The reason is simply that one has not found a convincing
and testable way to determine the flavor structure of massive neutrinos.
In this case one usually follows the bottom-up approach from a phenomenological
point of view, and a reconstruction of the effective Majorana neutrino mass matrix
$M^{}_\nu$ in terms of the neutrino masses and lepton flavor mixing parameters
as we have done belongs to this category. We are motivated by the question that
to what extent the six independent effective Majorana neutrino masses
$\vert{\langle m\rangle^{}_{\alpha\beta}}\vert$ (for $\alpha, \beta = e, \mu, \tau$)
can be constrained after all the neutrino oscillation parameters are well
measured. That is why we have mapped the 3D profiles of all the
$\vert{\langle m\rangle^{}_{\alpha\beta}}\vert$ with the help of current
neutrino oscillation data. Such a model-independent way can not only shed light
on the possible textures of $M^{}_\nu$ in the NMO or IMO case, but also reveal
their dependence on the absolute neutrino mass scale and one of the effective
Majorana CP phases.

It is the first time that the 3D mapping of all the six $\vert{\langle m\rangle^{}_{\alpha\beta}}\vert$ has been systematically studied in the present
work, although the 3D profile of $\vert{\langle m\rangle^{}_{ee}}\vert$ was
already discussed in the literature~\cite{Xing:2015zha,Xing:2016ymd,Cao:2019hli}.
Since a measurement of $\vert{\langle m\rangle^{}_{ee}}\vert$ itself in the
future $0\nu 2\beta$ experiments does not allow us to determine the Majorana
CP phases, it makes a lot of sense to look at the other five effective Majorana
neutrino masses no matter how difficult it is to measure them in reality.

When more accurate experimental data on all the neutrino oscillation parameters are
available in the near future, one may certainly follow the strategy and methods
outlined in this work to constrain the parameter space of each of
$\vert{\langle m\rangle^{}_{\alpha\beta}}\vert$ and unravel its phenomenological
implications in a more reliable way. The same methodology can also be
extended to the bottom-up study of possible flavor structures of massive neutrinos
beyond the standard three-flavor scheme (see, e.g., an example of this kind for
$\vert{\langle m\rangle^{}_{ee}}\vert$ in the (3+1) active-sterile neutrino mixing
scheme in Ref.~\cite{Liu:2017ago}).

\section*{Acknowledgements}

This work is supported in part by the National Natural
Science Foundation of China under grant No. 12075254, grant
No. 11775231 and grant No. 11835013.



\begin{thebibliography}{99}
\bibitem{Zyla:2020zbs}
P.~A.~Zyla \textit{et al.} [Particle Data Group],
``Review of Particle Physics,"
PTEP \textbf{2020} (2020) no.8, 083C01

\bibitem{Capozzi:2017ipn}
F.~Capozzi, E.~Di Valentino, E.~Lisi, A.~Marrone, A.~Melchiorri and A.~Palazzo,
``Global constraints on absolute neutrino masses and their ordering,''
Phys. Rev. D \textbf{101} (2020) 11, 116013
[arXiv:2003.08511 [hep-ph]].

\bibitem{Esteban:2020cvm}
I.~Esteban, M.~C.~Gonzalez-Garcia, M.~Maltoni, T.~Schwetz and A.~Zhou,
``The fate of hints: updated global analysis of three-flavor neutrino oscillations,''
JHEP \textbf{09} (2020), 178
[arXiv:2007.14792 [hep-ph]].

\bibitem{Abe:2019vii}
K.~Abe \textit{et al.} [T2K],
``Constraint on the matter\textendash{}antimatter symmetry-violating phase in neutrino oscillations,''
Nature \textbf{580} (2020) no.7803, 339-344
[erratum: Nature \textbf{583} (2020) no.7814, E16]
[arXiv:1910.03887 [hep-ex]].

\bibitem{Majorana:1937vz}
E.~Majorana,
``Teoria simmetrica dell\textquoteright{}elettrone e del positrone,''
Nuovo Cim. \textbf{14} (1937), 171-184

\bibitem{Pontecorvo:1957cp}
B.~Pontecorvo,
``Mesonium and anti-mesonium,"
Sov. Phys. JETP \textbf{6} (1957), 429

\bibitem{Schechter:1980gk}
J.~Schechter and J.~W.~F.~Valle,
``Neutrino Oscillation Thought Experiment,''
Phys. Rev. D \textbf{23} (1981), 1666

\bibitem{deGouvea:2002gf}
A.~de Gouvea, B.~Kayser and R.~N.~Mohapatra,
``Manifest CP Violation from Majorana Phases,''
Phys. Rev. D \textbf{67} (2003), 053004
[arXiv:hep-ph/0211394 [hep-ph]].

\bibitem{Xing:2013ty}
Z.~z.~Xing,
``Properties of CP Violation in Neutrino-Antineutrino Oscillations,''
Phys. Rev. D \textbf{87} (2013) no.5, 053019
[arXiv:1301.7654 [hep-ph]].

\bibitem{Rodejohann:2011mu}
W.~Rodejohann,
``Neutrino-less Double Beta Decay and Particle Physics,''
Int. J. Mod. Phys. E \textbf{20} (2011), 1833-1930
[arXiv:1106.1334 [hep-ph]].

\bibitem{Pontecorvo:1967fh}
B.~Pontecorvo,
``Neutrino Experiments and the Problem of Conservation of Leptonic Charge,"
Zh. Eksp. Teor. Fiz. \textbf{53} (1967), 1717-1725

\bibitem{Maki:1962mu}
Z.~Maki, M.~Nakagawa and S.~Sakata,
``Remarks on the unified model of elementary particles,"
Prog. Theor. Phys. \textbf{28} (1962), 870-880

\bibitem{Xing:2019vks}
Z.~z.~Xing,
``Flavor structures of charged fermions and massive neutrinos,''
Phys. Rept. \textbf{854} (2020), 1-147
[arXiv:1909.09610 [hep-ph]].

\bibitem{Feruglio:2021sir}
F.~Feruglio and A.~Romanino,
``Lepton flavor symmetries,''
Rev. Mod. Phys. \textbf{93} (2021) no.1, 015007
[arXiv:1912.06028 [hep-ph]].

\bibitem{Xing:2015zha}
Z.~z.~Xing, Z.~h.~Zhao and Y.~L.~Zhou,
``How to interpret a discovery or null result of the $0\nu 2\beta$ decay,''
Eur. Phys. J. C \textbf{75} (2015) no.9, 423
[arXiv:1504.05820 [hep-ph]].

\bibitem{Xing:2016ymd}
Z.~z.~Xing and Z.~h.~Zhao,
``The effective neutrino mass of neutrinoless double-beta decays: how possible to fall into a well,"
Eur. Phys. J. C \textbf{77} (2017) no.3, 192
[arXiv:1612.08538 [hep-ph]].

\bibitem{Cao:2019hli}
J.~Cao, G.~Y.~Huang, Y.~F.~Li, Y.~Wang, L.~J.~Wen, Z.~Z.~Xing, Z.~H.~Zhao and S.~Zhou,
``Towards the meV limit of the effective neutrino mass in neutrinoless double-beta decays,"
Chin. Phys. C \textbf{44} (2020) no.3, 031001
[arXiv:1908.08355 [hep-ph]].

\bibitem{Barger:2002vy}
V.~Barger, S.~L.~Glashow, P.~Langacker and D.~Marfatia,
``No go for detecting CP violation via neutrinoless double beta decay,''
Phys. Lett. B \textbf{540} (2002), 247-251
[arXiv:hep-ph/0205290 [hep-ph]].

\bibitem{Xing:2015fdg}
Z.~z.~Xing and Z.~h.~Zhao,
``A review of \ensuremath{\mu}-\ensuremath{\tau} flavor symmetry in neutrino physics,"
Rept. Prog. Phys. \textbf{79} (2016) no.7, 076201
[arXiv:1512.04207 [hep-ph]].

\bibitem{Dolinski:2019nrj}
M.~J.~Dolinski, A.~W.~P.~Poon and W.~Rodejohann,
``Neutrinoless Double-Beta Decay: Status and Prospects,"
Ann. Rev. Nucl. Part. Sci. \textbf{69} (2019), 219-251
[arXiv:1902.04097 [nucl-ex]].

\bibitem{Agostini:2018tnm}
M.~Agostini \textit{et al.} [GERDA],
``Improved Limit on Neutrinoless Double-beta Decay of $^{76}$Ge from GERDA Phase II,"
Phys. Rev. Lett. \textbf{120} (2018) no.13, 132503
[arXiv:1803.11100 [nucl-ex]].

\bibitem{Capozzi:2021fjo}
F.~Capozzi, E.~Di Valentino, E.~Lisi, A.~Marrone, A.~Melchiorri and A.~Palazzo,
``The unfinished fabric of the three neutrino paradigm,''
[arXiv:2107.00532 [hep-ph]].

\bibitem{Xing:2017cwb}
Z.~z.~Xing and J.~y.~Zhu,
``Neutrino mass ordering and \textbackslash{}mu-\textbackslash{}tau reflection symmetry breaking,''
Chin. Phys. C \textbf{41} (2017) no.12, 123103
[arXiv:1707.03676 [hep-ph]].

\bibitem{Frampton:2002yf}
P.~H.~Frampton, S.~L.~Glashow and D.~Marfatia,
``Zeroes of the neutrino mass matrix,''
Phys. Lett. B \textbf{536} (2002), 79-82
[arXiv:hep-ph/0201008 [hep-ph]].

\bibitem{Xing:2002ta}
Z.~z.~Xing,
``Texture zeros and Majorana phases of the neutrino mass matrix,''
Phys. Lett. B \textbf{530} (2002), 159-166
[arXiv:hep-ph/0201151 [hep-ph]].

\bibitem{Xing:2002ap}
Z.~z.~Xing,
``A Full determination of the neutrino mass spectrum from two zero textures of
the neutrino mass matrix,''
Phys. Lett. B \textbf{539} (2002), 85-90
[arXiv:hep-ph/0205032 [hep-ph]].

\bibitem{Fritzsch:2011qv}
H.~Fritzsch, Z.~z.~Xing and S.~Zhou,
``Two-zero Textures of the Majorana Neutrino Mass Matrix and Current Experimental Tests,''
JHEP \textbf{09} (2011), 083
[arXiv:1108.4534 [hep-ph]].

\bibitem{Zhou:2015qua}
S.~Zhou,
``Update on two-zero textures of the Majorana neutrino mass matrix in light of recent T2K, Super-Kamiokande and NO$\nu$A results,''
Chin. Phys. C \textbf{40} (2016) no.3, 033102
[arXiv:1509.05300 [hep-ph]].

\bibitem{Frigerio:2002rd}
M.~Frigerio and A.~Y.~Smirnov,
``Structure of neutrino mass matrix and CP violation,''
Nucl. Phys. B \textbf{640} (2002), 233-282
[arXiv:hep-ph/0202247 [hep-ph]].

\bibitem{Merle:2006du}
A.~Merle and W.~Rodejohann,
``The Elements of the neutrino mass matrix: Allowed ranges and implications of texture zeros,''
Phys. Rev. D \textbf{73} (2006), 073012
[arXiv:hep-ph/0603111 [hep-ph]].

\bibitem{Grimus:2012ii}
W.~Grimus and P.~O.~Ludl,
``Correlations of the elements of the neutrino mass matrix,''
JHEP \textbf{12} (2012), 117
[arXiv:1209.2601 [hep-ph]].

\bibitem{Ohlsson:2013xva}
T.~Ohlsson and S.~Zhou,
``Renormalization group running of neutrino parameters,''
Nature Commun. \textbf{5} (2014), 5153
[arXiv:1311.3846 [hep-ph]].

\bibitem{Witten:2000dt}
E.~Witten,
``Lepton number and neutrino masses,''
Nucl. Phys. B Proc. Suppl. \textbf{91} (2001), 3-8
[arXiv:hep-ph/0006332 [hep-ph]].

\bibitem{Liu:2017ago}
J.~H.~Liu and S.~Zhou,
``Another look at the impact of an eV-mass sterile neutrino on the effective neutrino mass of neutrinoless double-beta decays,''
Int. J. Mod. Phys. A \textbf{33} (2018) no.02, 1850014
[arXiv:1710.10359 [hep-ph]].

\end{thebibliography}
\end{document}